\documentclass[sn-mathphys-num]{sn-jnl}% Math and Physical Sciences Numbered Reference Style 
\usepackage{graphicx}%
\usepackage{multirow}%
\usepackage{amsmath,amssymb,amsfonts}%
\usepackage{amsthm}%
\usepackage{mathrsfs}%
\usepackage[title]{appendix}%
\usepackage{xcolor}%
\usepackage{textcomp}%
\usepackage{manyfoot}%
\usepackage{booktabs}%
\usepackage{algorithm}%
\usepackage{algorithmicx}%
\usepackage[noend]{algpseudocode}%
\usepackage{listings}%

\usepackage{fancyvrb} 
\usepackage[utf8]{inputenc}
\usepackage{hyperref} 
\usepackage{array}
\usepackage{dcolumn}
\usepackage{paralist}
\usepackage{fullwidth} 
\usepackage{subcaption}
\usepackage{scalerel}
\usepackage{mathtools}
\usepackage{siunitx}
\usepackage{tikz, pgfplots, pgfplotstable}
\usepgfplotslibrary{groupplots}
\usetikzlibrary{shapes}
\usetikzlibrary{calc}
\pgfplotsset{compat=newest}
\usetikzlibrary{pgfplots.statistics} %Tikz
\newcommand{\nR}{r}
\newcommand{\nRMin}{\nR^{min}}
\newcommand{\nRMax}{\nR^{max}}
\newcommand{\nC}{C}
\newcommand{\nCMin}{\nC^{min}}
\newcommand{\nCMax}{\nC^{max}}
\newcommand{\nD}{d}
\newcommand{\nPr}{p}
\newcommand{\nPe}{\tau}

\newcommand{\nEtTask}[1][]{\mathcal{T}_{#1}}
\newcommand{\nEtJobS}{J}
\newcommand{\nEtJob}[2]{\nEtJobS_{#1,#2}}

\newcommand{\nRMinEtTask}[1][]{\nRMin_{#1}}
\newcommand{\nRMaxEtTask}[1][]{\nRMax_{#1}}
\newcommand{\nCMinEtTask}[1][]{\nCMin_{#1}}
\newcommand{\nCMaxEtTask}[1][]{\nCMax_{#1}}
\newcommand{\nDEtTask}[1][]{\nD_{#1}}
\newcommand{\nPrEtTask}[1][]{\nPr_{#1}}
\newcommand{\nPeEtTask}[1][]{\nPe_{#1}}

\newcommand{\nRMinEtJob}[2]{\nRMin_{#1,#2}}
\newcommand{\nRMaxEtJob}[2]{\nRMax_{#1,#2}}
\newcommand{\nREtJob}[2]{\nR_{#1,#2}}
\newcommand{\nCMinEtJob}[2]{\nCMin_{#1,#2}}
\newcommand{\nCMaxEtJob}[2]{\nCMax_{#1,#2}}
\newcommand{\nCEtJob}[2]{\nC_{#1,#2}}
\newcommand{\nDEtJob}[2]{\nD_{#1,#2}}
\newcommand{\nPrEtJob}[2]{\nPr_{#1,#2}}

\newcommand{\nHyperPe}{H}
\newcommand{\myInf}{\infty}
\newcommand{\nAppJobs}{\nEtJobS^A}
\newcommand{\nPolicy}{\Pi}

\newcommand{\nEtTaskRMin}[1][]{\nEtTask[#1].\nRMin}
\newcommand{\nEtTaskRMax}[1][]{\nEtTask[#1].\nRMax}
\newcommand{\nEtTaskCMin}[1][]{\nEtTask[#1].\nCMin}
\newcommand{\nEtTaskCMax}[1][]{\nEtTask[#1].\nCMax}

\newcommand{\nEtTaskPe}[1][]{\nEtTask[#1].\nPe}

\newcommand{\nEtJobRMin}[2]{\nEtJob{#1}{#2}.\nRMin}
\newcommand{\nEtJobRMax}[2]{\nEtJob{#1}{#2}.\nRMax}

\newcommand{\nEtJobCMin}[2]{\nEtJob{#1}{#2}.\nCMin}
\newcommand{\nEtJobCMax}[2]{\nEtJob{#1}{#2}.\nCMax}

\newcommand{\nEtJobD}[2]{\nEtJob{#1}{#2}.\nD}

\newcommand{\nVert}[1][]{v_{#1}}

\newcommand{\nEdge}[1][]{a_{#1}}
\newcommand{\nEft}[1][]{e_{#1}}
\newcommand{\nLft}[1][]{l_{#1}}
\newcommand{\nInt}[1][]{[\nEft[#1],\nLft[#1]]}
\newcommand{\nVertLvl}[1][]{L_{#1}}
\newcommand{\nVertLvlN}[1][]{V^{#1}}
\newcommand{\nVertLvlExp}[1][]{\nVertLvl[#1]^{ex}}

\newcommand{\nEdgePath}[1][]{\nEtJobS^{v_{#1}}}

\newcommand{\nCeJob}[1][]{\nEtJobS^{#1}_{ce}}
\newcommand{\nPeJob}[1][]{\nEtJobS^{#1}_{pe}}
\newcommand{\nLftExt}[1][]{{\nLft}^{ext}}

\newcommand{\nCritJob}{\nEtJobS^c}
\newcommand{\nCritTime}{t^c}

\newcommand{\nVertEft}[1][]{\nVert[#1].\nEft}
\newcommand{\nVertLft}[1][]{\nVert[#1].\nLft}
\newcommand{\nVertIn}[1][]{\nVert[#1].in}
\newcommand{\nVertOut}[1][]{\nVert[#1].out}
\newcommand{\nVertAppJobs}[1][]{\nVert[#1].\nAppJobs}

\newcommand{\nVertCritTime}[1][]{\nVert[#1].\nCritTime}
\newcommand{\nVertCritJob}[1][]{\nVert[#1].\nCritJob}
\newcommand{\nVertInt}[1][]{[\nVert[#1].\nEft, \nVert[#1].\nLft]}

\newcommand{\nGen}{A}

\newcommand{\nGenRandPerc}{\nGen^P}
\newcommand{\nGenPercJitt}{\nGenRandPerc_j}
\newcommand{\nGenPercCVar}{\nGenRandPerc_c}

\newcommand{\nDataset}[2]{\mathcal{D}_{#1}^{#2}}

\definecolor{Red}{HTML}{bd0026}
\definecolor{RedLight}{HTML}{f03b20}
\definecolor{Gray}{HTML}{969696}
\definecolor{GrayLight}{HTML}{cccccc}
\definecolor{Green}{HTML}{74c476}

\newcommand{\job}[4] { % x1, x2, y, name
		    \path[fill=RedLight] (axis cs: #1+0.02, #3+0.2) rectangle (axis cs: #2-0.02, #3+0.8);
			\node (a#1-#3) at (axis cs: #1,#3+0.5) {};
			\node (b#2-#3) at (axis cs: #2, #3+0.5) {};	
			\node[font=\scriptsize] at ($(a#1-#3)!0.5!(b#2-#3)$)  {$#4$};
}
\newcommand{\jobjitter}[3] { % x1, x2, y
		    \path[fill=Red] (axis cs: #1-0.02, #3+0.2) rectangle (axis cs: #2, #3+0.8);
}
\newcommand{\pozadisvetle}[3] { % x1, x2, y
		    \path[fill=GrayLight] (axis cs: #1, #3) rectangle (axis cs: #2, #3+1);
}
\newcommand{\pozaditmave}[3] { % x1, x2, y
		    \path[fill=Gray] (axis cs: #1, #3) rectangle (axis cs: #2, #3+1);
}
\newcommand{\deadline}[2] { % x1, y
		    \path[fill=Green] (axis cs: #1-0.03, #2) rectangle (axis cs: #1+0.03, #2+1);
}

\theoremstyle{thmstyleone}%
\theoremstyle{thmstyletwo}%

\theoremstyle{thmstylethree}%

\begin{document}

\title[Revisiting the Schedule Graph Generation for the Exact and Sustainable Analysis of Non-preemptive Scheduling]{Revisiting the Schedule Graph Generation for the Exact and Sustainable Analysis of Non-preemptive Scheduling}

\author*[1]{\fnm{Marek} \sur{Vlk}}\email{vlkmarek8@gmail.com}

\author[2]{\fnm{Marek} \sur{Jaroš}}\email{jarosm11@fel.cvut.cz}

\author[1]{\fnm{Zdeněk} \sur{Hanzálek}}\email{zdenek.hanzalek@cvut.cz}

\affil[1]{\orgdiv{Czech Institute of Informatics, Robotics and Cybernetics}, \orgname{Czech Technical University in Prague}, \orgaddress{\street{Jugoslávských partyzánů 1580/3}, \city{Prague}, \postcode{160 00}, \country{Czech Republic}}}

\affil[2]{\orgdiv{Faculty of Electrical Engineering}, \orgname{Czech Technical University in Prague}, \orgaddress{\street{Technická~2}, \city{Prague}, \postcode{166 27},  \country{Czech Republic}}}

\abstract{This paper addresses the problem of scheduling non-preemptive tasks with release jitter and execution time variation on a uniprocessor. We show that the schedulability analysis based on schedule graph generation, proposed by Nasri and Brandenburg [RTSS 2017], produces negative results when it could be easily avoided by slightly reformalizing the notion of non-work-conserving policies. In this work, we develop a schedulability analysis that constructs the schedule graph using new job-eligibility rules and is exact and sustainable for both work-conserving and enhanced formalization of non-work-conserving policies. Besides, the experimental evaluation shows that our schedulability analysis is substantially faster.}




\sloppy

\maketitle

%\end{document}

\section{Introduction}

Ensuring the temporal correctness of safety-critical real-time systems is a fundamental challenge, often addressed through response-time analysis (RTA). The primary objective of RTA is to ascertain the worst-case response time (WCRT) for a set of jobs scheduled by a given policy on computing resources. A real-time system is schedulable if the WCRT of each job does not exceed its deadline. However, the complexity of the RTA problem escalates, particularly for periodic tasks scheduled by a job-level fixed-priority policy, even on a uniprocessor platform, rendering most variations of the problem $\mathcal{NP}$-hard \cite{ekberg2020rate} or co-$\mathcal{NP}$-hard \cite{baruah2004scheduling}. This computational intractability not only poses formidable theoretical challenges but also hampers the practical application of these scheduling policies in real-world safety-critical systems. As such, there is a pressing need to delve into innovative approaches that not only address the inherent complexity but also pave the way for more efficient and scalable analyses, thereby facilitating the dependable design and implementation of safety-critical real-time systems.

While there are efficient algorithms for scheduling preemptive tasks \cite{park2004feasibility}, the difficulty of scheduling non-preemptive tasks with release jitter and execution time variation lies in the fact that it may exhibit anomalies, i.e., the scenario deemed worst-case does not result in a deadline miss, but a different scenario does. A schedulability analysis is sustainable if any problem instance deemed schedulable by the analysis is schedulable not only for the worst-case but for all conceivable scenarios \cite{SUS}.

\subsection{Related Work}

The conventional methods employed for sustainable schedulability analyses involve a pessimistic over-approximation of the true worst-case scenario with contrived situations that may be unrealistic at runtime. For instance, in the traditional RTA applied to non-preemptive fixed-priority scheduling \cite{davis2007controller, tindell1994extendible}, the bound on the response time of a task is derived assuming that, in the worst case, a job experiences maximal blocking from lower-priority jobs and maximal interference from higher-priority jobs, which may be overly pessimistic for periodic tasks.

On the contrary, the potential for exact and sustainable analysis results is offered by alternative approaches that are based on model checking (e.g., \cite{guan2007exact,sun2016pre}) or exhaustive exploration of all system states \cite{burmyakov2015exact,baker2007brute,bonifaci2012feasibility}. Nonetheless, these approaches grapple with significant state-space explosion issues, making them impractical even for small problem instances.

Recently, a new branch of research emerged, centered around the concept of the schedule graph \cite{SANS}. The schedule graph (or schedule abstraction graph) abstracts the space of available decisions that can be taken by a given scheduler when dispatching a set of jobs onto a resource. This exploration of decision space involves constructing a combinatorial graph where nodes correspond to the state of the resource after the execution of a set of jobs, and whose edges correspond to possible scheduling decisions (which is to determine a job to be processed next) that shape the transition of the system states. Throughout the construction of the graph, the response time of each job dispatched along an edge is examined. Upon completing the graph, the method thus yields the minimum and maximum response time for each job across all scenarios.

Since their seminal work on a uniprocessor \cite{SANS}, the notion of the schedule graph has been extended to global scheduling (multiple processors) \cite{nasri2018response} and global scheduling with precedences \cite{nasri2019response,nogd2020response}. The methodology has been improved by a so-called partial order reduction, which is applicable to the problems both with \cite{ranjha2022partial} and without \cite{ranjha2023partial} precedences.

A key strength of the schedule graph lies in its ability to accommodate non-work-conserving policies, known for their significant schedulability advantages \cite{PRM,nasri2016non}. However, \cite{SANS} is the only work addressing the non-work-conserving policies using the schedule graph. The authors devised a schedule graph generation algorithm that is exact and sustainable for both work-conserving and non-work-conserving policies. Despite these claims, our investigation has revealed instances where the schedulability analysis used with non-work-conserving policies produces negative results when it could be easily avoided by slightly reformalizing the notion of non-work-conserving policies.

\subsection{Outline and Contributions of this Paper}

We revisit the schedulability analysis for non-preemptive tasks on a uniprocessor proposed in \cite{SANS}. We refer to the approach presented in \cite{SANS} as Schedule Graph Analysis with Single Eligibility (SGA-SE). We extend the ideas of SGA-SE and enhance the concept of non-work-conserving policies such that the set of schedulable instances is enlarged. Our new algorithm will be distinguished by being referred to as Schedule Graph Analysis with Multiple Eligibility (SGA-ME).

The contributions of this paper are:
\begin{itemize}
    \item discovered easily avoidable pessimism in the schedulability analysis for non-work-conserving policies presented in \cite{SANS},
    \item definitions of certainly-eligible and possibly-eligible jobs that lead to a new algorithm for the schedulability analysis that is exact and sustainable for both work-conserving and enhanced formalization of non-work-conserving policies,
    \item the new algorithm implementation that is by an order of magnitude faster than \cite{SANS}.
\end{itemize}

To maximize the transparency and reproducibility of our results, all the source codes of our implementation and the benchmarking instances are publicly available\footnote{\url{https://github.com/redakez/ettt-scheduling}}.

The rest of the paper is organized as follows. Section~\ref{2_formal_desc} formally describes the problem. Section~\ref{3_ET_solutions} focuses on the schedulability analysis whose performance is evaluated in Section~\ref{4_ET_benchmark}. The paper concludes in Section~\ref{7_Conclusion}.

\section{Problem Description} \label{2_formal_desc}

This section gives a formal description of the problem.
All variables defined in Section \ref{2_formal_desc} are integers. All proposed algorithms in later sections also work with integer values only.

\subsection{Definition of Tasks and Jobs}

The problem is given by a set of tasks $\nEtTask = (\nEtTask[1],\dots,\nEtTask[n])$, where each task $\nEtTask[i]$ is specified by period $\nPeEtTask[i]$, the earliest release time $\nRMinEtTask[i]$, the latest release time $\nRMaxEtTask[i] \geq \nRMinEtTask[i]$, best-case execution time $\nCMinEtTask[i]$, worst-case execution time $\nCMaxEtTask[i] \geq \nCMinEtTask[i]$, deadline $\nDEtTask[i]$ and priority $\nPrEtTask[i] \geq 0$. 

Each task $\nEtTask[i]$ consists of a set of jobs $\nEtJob{i}{1},\dots,\nEtJob{i}{h_i}$, where $\nEtJob{i}{j}$ is the j-th job of task $\nEtTask[i]$ and $h_i$ is the last job of task $\nEtTask[i]$ in the \textit{observation interval}. Each job $\nEtJob{i}{j}$ is specified by the following parameters that are inferred from task $\nEtTask[i]$ as follows. The priority $\nPrEtJob{i}{j} = \nPrEtTask[i]$, the earliest release time $\nRMinEtJob{i}{j} = \nRMinEtTask[i] + (j-1) \cdot \nPeEtTask[i]$, the latest release time $\nRMaxEtJob{i}{j} = \nRMaxEtTask[i] + (j-1) \cdot \nPeEtTask[i]$, best-case execution time $\nCMinEtJob{i}{j} = \nCMinEtTask[i]$, worst-case execution time $\nCMaxEtJob{i}{j} = \nCMaxEtTask[i]$ and deadline $\nDEtJob{i}{j} = \nDEtTask[i] + (j-1) \cdot \nPeEtTask[i]$.

In practice, tasks generate infinite sequences of jobs, that is why we need to work with the observation interval, which is the representative timeframe in which the release pattern of the set of tasks $\nEtTask$ repeats. It must satisfy the requirement that if the set of tasks is proved to be schedulable on the observation interval, then the set of tasks is also schedulable on the infinite horizon. Since the work we are revising \cite{SANS} is described technically for a non-periodic finite set of jobs and since finding the smallest possible observation interval is a very complex problem, it stays out of the scope of this paper.

\subsection{Scheduling of Jobs}\label{defea}

The base of the problem is online scheduling of a set of jobs defined by a set of tasks $\nEtTask = (\nEtTask[1],\dots,\nEtTask[n])$ on a uniprocessor. 

During run time, each job $\nEtJob{i}{j}$ is released at a priori unknown time $\nREtJob{i}{j} \in [\nRMinEtJob{i}{j}, \nRMaxEtJob{i}{j}]$. This behavior is referred to as \textit{release jitter}. Once job $\nEtJob{i}{j}$ is released, the value $\nREtJob{i}{j}$ becomes known. If job $\nEtJob{i}{j}$ is started at time $t$, then it occupies the processor during interval $[t, t + \nCEtJob{i}{j})$ where $\nCEtJob{i}{j} \in [\nCMinEtJob{i}{j}, \nCMaxEtJob{i}{j}]$ is a priori unknown execution time. This behavior is referred to as \textit{execution time variation}. %Time $t$ is referred to as \textit{time of execution}. 
If job $\nEtJob{i}{j}$ finishes execution at time $t + \nCEtJob{i}{j} > \nDEtJob{i}{j}$, then the online scheduler yields a \textit{deadline miss}.

Additionally, $\nEtJob{i}{j}$ is \textit{finished} if it has been executed, i.e., it was picked by the online scheduler and then occupied the processor for $\nCEtJob{i}{j}$ units of time. A job is \textit{unfinished} if it is not finished yet. A job $\nEtJob{i}{j}$ is \textit{certainly released} at time $t$ if $\nRMaxEtJob{i}{j} \leq t$ and \textit{possibly released} at time $t$ if $\nRMinEtJob{i}{j} \leq t < \nRMaxEtJob{i}{j}$.

We define a set of \textit{applicable jobs} as a set of jobs that contains the first unfinished job of each task, i.e., a set of applicable jobs $\nAppJobs$ contains jobs $\nEtJob{i}{j}$ that satisfy: $\nEtJob{i}{j}$ is unfinished $\wedge \; (j=1 ~\vee~ \nEtJob{i}{j-1}$ is finished$)$. Notice that the set of applicable jobs is calculated from the set of finished jobs. Also note that the set of applicable jobs may contain at most $n$ jobs, one from each task, and may also be empty, which happens when all jobs are finished.

\subsection{Scheduling Policies} \label{2_3_policies}

Scheduling policy is a function $\nPolicy(t,\nAppJobs)$ which for a given set of applicable jobs $\nAppJobs$ and time $t$ returns a job that should be scheduled next. The policy can also return $null$, which can happen if the set $\nAppJobs$ is empty or if the non-work conserving policy chooses to idle until some unreleased job releases.

It is assumed that the scheduling overhead resulting from the run time of an online scheduler executing a scheduling policy is insignificant and is therefore ignored. The priority value $\nPrEtTask$ of tasks and jobs is used only in some scheduling policies. Tasks with the numerically smallest value of $\nPrEtTask$ have the highest priority.

We will work with the following scheduling policies. The Earliest Deadline First (EDF) picks the job with the earliest deadline, i.e., smallest $\nDEtJob{i}{j}$, out of all released applicable jobs. Fixed Priority - EDF (FP-EDF) picks the job with the highest priority, breaking ties by taking the job with the earliest deadline. Precautious FP-EDF (P-FP-EDF) works like FP-EDF but avoids picking a job with a priority lower than the highest one if it may cause a deadline miss for some applicable job of the highest priority. It is assumed that all tasks have unique identification numbers which are used in scheduling policies to break ties to keep the policies deterministic. The scheduling policies are elaborated in more detail in \ref{sec:policies}. 

\subsection{Execution Scenarios}

For a set of tasks $\nEtTask = (\nEtTask[1],\dots,\nEtTask[n])$, an \textit{execution scenario} $\gamma = (C,R)$ is a set of execution times $C = (C_1,\dots,C_n)$ and release times $R = (R_1,\dots,R_n)$, where $C_i = (\nCEtJob{i}{1},\dots,\nCEtJob{i}{h_i})$, $R_i = (\nREtJob{i}{1},\dots,\nREtJob{i}{h_i})$, $\nCEtJob{i}{j} \in [\nCMinEtJob{i}{j}, \nCMaxEtJob{i}{j}]$ and $\nREtJob{i}{j} \in [\nRMinEtJob{i}{j}, \nRMaxEtJob{i}{j}]$. In other words, an execution scenario specifies the release time and execution time for every job of every task in $\nEtTask$. The values of $\nREtJob{i}{j}$ and $\nCEtJob{i}{j}$ are unknown a priori and are revealed during the run time execution. Note that for a given execution scenario $\gamma$ and policy $\nPolicy$, the times when the jobs start their execution are deterministic.

Given policy $\nPolicy$, a set of tasks $\nEtTask$\ is \textit{schedulable} if there exists no execution scenario resulting in a deadline miss for an online scheduler that uses policy $\nPolicy$. On the contrary, a set of tasks $\nEtTask$\ is \textit{non-schedulable} if some execution scenario results in a deadline miss.

\subsection{Illustrative Example}

To give a better understanding of the concept, we present the following illustrative problem instance. 
Let $\nEtTask = (\nEtTask[1],\nEtTask[2],\nEtTask[3])$. The parameters of these tasks can be seen in Table \ref{table:ex1_table}. The goal is to determine whether the tasks are schedulable under the EDF scheduling policy. Note that EDF does not make use of priority values $\nPrEtTask$ and they are therefore undefined.

\setlength\extrarowheight{2pt}
\begin{table}
    \caption{Parameters of tasks $\nEtTask = (\nEtTask[1],\nEtTask[2],\nEtTask[3])$}
    \label{table:ex1_table}
    \centering
    \begin{tabular}{c c c c c c c} 
    \toprule
    & $\nRMinEtTask$ & $\nRMaxEtTask$ & $\nCMinEtTask$ & $\nCMaxEtTask$ & $\nDEtTask$ & $\nPeEtTask$ \\
    \midrule
    $\nEtTask[1]$ & 2 & 5 & 5 & 7 & 16 & 20 \\
    $\nEtTask[2]$ & 1 & 1 & 2 & 4 & 8 & 10 \\
    $\nEtTask[3]$ & 0 & 0 & 1 & 1 & 5 & 5 \\
    \bottomrule
    \end{tabular}
\end{table}
\setlength\extrarowheight{0pt}

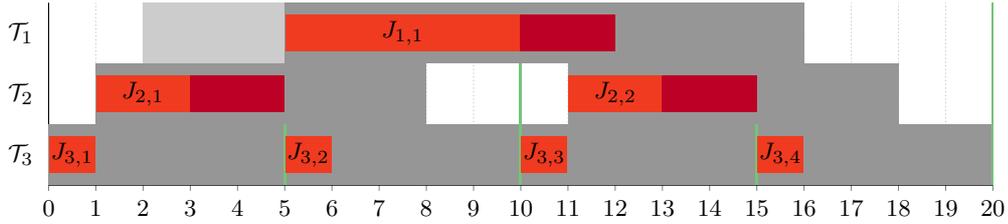
\begin{figure}[!hbt]
\centering
\begin{tikzpicture}
\pgfmathsetlengthmacro\MajorTickLength{
      \pgfkeysvalueof{/pgfplots/major tick length} * 0.5}
		\begin{axis}[
	        width=14cm,
	        height=4cm,
	        xmajorgrids={true},
			major grid style={densely dotted},
			ytick style={draw=none},
            axis lines = left,
            axis line style=-,
			enlarge x limits=0,
			enlarge y limits=0,
			tick align=outside,
			major tick length=\MajorTickLength,
			ymin=0,
			ymax=3,
			xmin=0,
			xmax=20,
			ytick={0.5,1.5,2.5},
			xtick={0,1,2,...,20},
			yticklabels={$\nEtTask[3]$,$\nEtTask[2]$,$\nEtTask[1]$},
			x tick label style={font=\scriptsize},
			y tick label style={font=\small}]  
   \pozadisvetle{2}{5}{2}
   \pozaditmave{5}{16}{2}
   \pozaditmave{1}{8}{1}
   \pozaditmave{11}{18}{1}
   \pozaditmave{0}{20}{0}
   \deadline{20}{2}
   \deadline{10}{1}
   \deadline{20}{1}
   \deadline{5}{0}
   \deadline{10}{0}
   \deadline{15}{0}
   \deadline{20}{0}
   \jobjitter{10}{12}{2}
   \jobjitter{3}{5}{1}
   \jobjitter{13}{15}{1}
   \job{5}{10}{2}{\nEtJob{1}{1}}
   \job{1}{3}{1}{\nEtJob{2}{1}}
   \job{11}{13}{1}{\nEtJob{2}{2}}
   \job{0}{1}{0}{\nEtJob{3}{1}}
   \job{5}{6}{0}{\nEtJob{3}{2}}
   \job{10}{11}{0}{\nEtJob{3}{3}}
   \job{15}{16}{0}{\nEtJob{3}{4}}
\end{axis}
\end{tikzpicture}
    \caption{A loose chart of the example instance given in Table~\ref{table:ex1_table}. Each row represents a different task. The top row represents task $\nEtTask[1]$, the middle row represents task $\nEtTask[2]$ and the bottom row represents task $\nEtTask[3]$.}
    \label{fig:ex1_loose}
\end{figure}

Choosing the observation interval as 20 implies that tasks $\nEtTask[1]$,$\nEtTask[2]$ and $\nEtTask[3]$ will have 1, 2 and 4 jobs respectively. Individual jobs are denoted as $\nEtTask[1] = (\nEtJob{1}{1})$, $\nEtTask[2] = (\nEtJob{2}{1},\nEtJob{2}{2})$ and $\nEtTask[3] = (\nEtJob{3}{1},\nEtJob{3}{2},\nEtJob{3}{3},\nEtJob{3}{4})$.

The instance is visualized in \figureautorefname{}~\ref{fig:ex1_loose} but bear in mind that this visualization displays only defined properties of tasks $\nEtTask$, does not care about non-confliting execution on monoprocessor, and says nothing about the used policy or when a job is executed. This is due to the fact that in our case, the time of execution of each job varies based on an execution scenario. Despite this, the visualization still provides some insight into the instance. We call this type of visualization a \textit{loose chart}. Because the time of execution of each job is unknown a priori, each job is placed at its maximal release time.

In the loose charts, gray areas show when a job can be executed. Release jitter is displayed using a bright gray color. Minimal and maximal execution times are displayed using bright and dark red colors. Green lines denote periods.

Given an execution scenario and a scheduling policy, the instance can be visualized using a chart that we call a \textit{regular chart}. A regular chart is a Gantt chart with gray areas showing when a job can be executed. Consider the EDF scheduling policy and the execution scenario in which all jobs exhibit the worst-case execution times and latest release times. That is:
$$\nCEtJob{1}{1} = 7, \nREtJob{1}{1} = 5$$ 
$$\nCEtJob{2}{1} = \nCEtJob{2}{2} = 4, \nREtJob{2}{1} = 1, \nREtJob{2}{2} = 11$$
$$\nCEtJob{3}{1} = \nCEtJob{3}{2} = \nCEtJob{3}{3} = \nCEtJob{3}{4} = 1$$
$$\nREtJob{3}{1} = 0, \nREtJob{3}{2} = 5, \nREtJob{3}{3} = 10, \nREtJob{3}{4} = 15$$

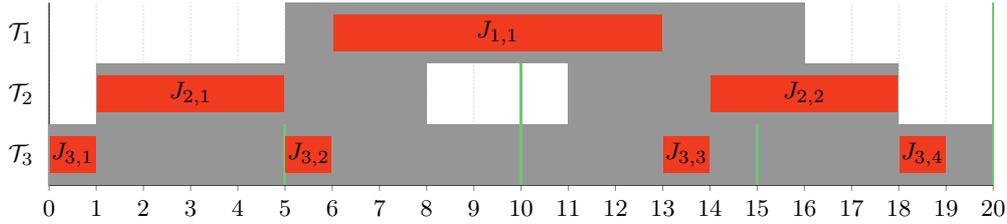
\begin{figure}[!hbt]
\centering
\begin{tikzpicture}
\pgfmathsetlengthmacro\MajorTickLength{
      \pgfkeysvalueof{/pgfplots/major tick length} * 0.5}
		\begin{axis}[
	        width=14cm,
	        height=4cm,
	        xmajorgrids={true},
			major grid style={densely dotted},
			ytick style={draw=none},
            axis lines = left,
            axis line style=-,
			enlarge x limits=0,
			enlarge y limits=0,
			tick align=outside,
			major tick length=\MajorTickLength,
			ymin=0,
			ymax=3,
			xmin=0,
			xmax=20,
			ytick={0.5,1.5,2.5},
			xtick={0,1,2,...,20},
			yticklabels={$\nEtTask[3]$,$\nEtTask[2]$,$\nEtTask[1]$},
			x tick label style={font=\scriptsize},
			y tick label style={font=\small}]
   \pozaditmave{5}{16}{2}
   \pozaditmave{1}{8}{1}
   \pozaditmave{11}{18}{1}
   \pozaditmave{0}{20}{0}
   \deadline{20}{2}
   \deadline{10}{1}
   \deadline{20}{1}
   \deadline{5}{0}
   \deadline{10}{0}
   \deadline{15}{0}
   \deadline{20}{0}
   \job{6}{13}{2}{\nEtJob{1}{1}}
   \job{1}{5}{1}{\nEtJob{2}{1}}
   \job{14}{18}{1}{\nEtJob{2}{2}}
   \job{0}{1}{0}{\nEtJob{3}{1}}
   \job{5}{6}{0}{\nEtJob{3}{2}}
   \job{13}{14}{0}{\nEtJob{3}{3}}
   \job{18}{19}{0}{\nEtJob{3}{4}}
\end{axis}
\end{tikzpicture}
    \caption{A regular chart of the example instance given in Table~\ref{table:ex1_table} assuming the worst-case execution times and latest release times.}
    \label{fig:ex1_WC}
\end{figure}

A regular chart using this scenario and the EDF scheduling policy is visualized in \figureautorefname{}~\ref{fig:ex1_WC}. This execution scenario does not result in a deadline miss. However, it does not mean that the set of tasks $\nEtTask = (\nEtTask[1],\nEtTask[2],\nEtTask[3])$ is schedulable.

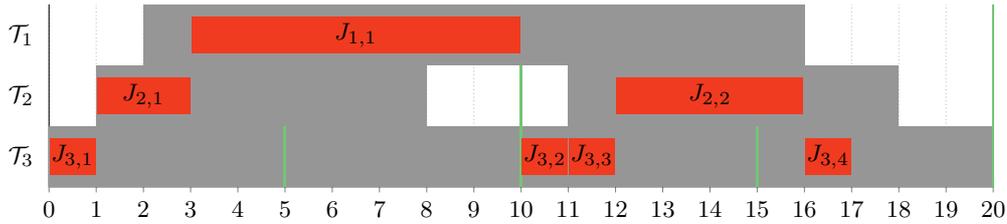
\begin{figure}[!hbt]
\centering
\begin{tikzpicture}
\pgfmathsetlengthmacro\MajorTickLength{
      \pgfkeysvalueof{/pgfplots/major tick length} * 0.5}
		\begin{axis}[
	        width=14cm,
	        height=4cm,
	        xmajorgrids={true},
			major grid style={densely dotted},
			ytick style={draw=none},
            axis lines = left,
            axis line style=-,
			enlarge x limits=0,
			enlarge y limits=0,
			tick align=outside,
			major tick length=\MajorTickLength,
			ymin=0,
			ymax=3,
			xmin=0,
			xmax=20,
			ytick={0.5,1.5,2.5},
			xtick={0,1,2,...,20},
			yticklabels={$\nEtTask[3]$,$\nEtTask[2]$,$\nEtTask[1]$},
			x tick label style={font=\scriptsize},
			y tick label style={font=\small}]
   \pozaditmave{2}{16}{2}
   \pozaditmave{1}{8}{1}
   \pozaditmave{11}{18}{1}
   \pozaditmave{0}{20}{0}
   \deadline{20}{2}
   \deadline{10}{1}
   \deadline{20}{1}
   \deadline{5}{0}
   \deadline{10}{0}
   \deadline{15}{0}
   \deadline{20}{0}
   \job{3}{10}{2}{\nEtJob{1}{1}}
   \job{1}{3}{1}{\nEtJob{2}{1}}
   \job{12}{16}{1}{\nEtJob{2}{2}}
   \job{0}{1}{0}{\nEtJob{3}{1}}
   \job{10}{11}{0}{\nEtJob{3}{2}}
   \job{11}{12}{0}{\nEtJob{3}{3}}
   \job{16}{17}{0}{\nEtJob{3}{4}}
\end{axis}
\end{tikzpicture}
    \caption{A regular chart of the example instance given in Table~\ref{table:ex1_table} with execution scenario that results in a deadline miss.}
    \label{fig:ex1_miss}
\end{figure}

Consider another execution scenario such that job $\nEtJob{1}{1}$ releases at $\nREtJob{1}{1} = \nRMinEtJob{1}{1}$ instead of $\nRMaxEtJob{1}{1}$ and execution time of job $\nEtJob{2}{1}$ is $\nCEtJob{2}{1} = \nCMinEtJob{2}{1}$ instead of $\nCMaxEtJob{2}{1}$. This execution scenario results in a deadline miss for job $\nEtJob{3}{2}$ at time 10 as can be seen in \figureautorefname{}~\ref{fig:ex1_miss}.

This so-called scheduling anomaly shows that if we want to make the schedulability analysis sustainable, it does not suffice to consider only the worst-case execution times and latest release times when analyzing schedulability.

\section{Schedulability Analysis}  \label{3_ET_solutions}

In this section, we propose a new algorithm for schedulability analysis, which is an extension of \cite{SANS} such that it is exact and sustainable also for enhanced formalization of non-work-conserving policies. We refer to our new algorithm as SGA-ME. SGA-ME also leverages the schedule graph generation, but what is completely different are the eligibility rules for adding new vertices in the expansion phase. Also, thanks to these new definitions of eligibility rules, SGA-ME is significantly faster than SGA-SE.

First, we focus on the basics of the schedule graph. We then provide a rough description of the generation algorithm and a step-by-step illustrative example of graph generation. The rules for schedule graph generation are described in detail afterward.

\subsection{Schedule Graph}

The schedule graph is a directed acyclic graph, in which each arc corresponds to the execution of some job and each vertex $\nVert[k]$ corresponds to a set of finished jobs (determined by the arcs on the path from the root vertex to $\nVert[k]$). Besides, each vertex $\nVert[k]$ contains a range of times $\nInt[k]$, during which the last job in the set of finished jobs may finish.
In other words, the processor may become available at any time point from the interval $\nInt[k]$.

The graph starts with root vertex $\nVert[0]$ with no finished jobs and a range of times $[0,0]$. The schedule graph is gradually built from the root vertex up to a graph that encapsulates all possible sequences of job executions.

Formally, the schedule graph $\mathcal{G} = (V,A)$ is defined as a directed acyclic graph, where each vertex $\nVert[k]$ has a label consisting of its ID and \textit{finish time interval} $\nInt[k]$, where $\nEft[k]$ and $\nLft[k]$ are the earliest finish time and the latest finish time, respectively. Furthermore, each arc $\nEdge \in A$ has a label corresponding to a single job that is to be scheduled next. Note that there may be several vertices with the same set of finished jobs and multiple arcs can have the same job label.

It is useful to see the schedule graph as a directed level-structured graph \cite{LEVEL}. This means $V$ can be split into disjoint sets based on the distance from root vertex $\nVert[0]$. Here, the distance of vertex $\nVert[k]$ from root vertex $\nVert[0]$ is the length of the path with the smallest number of arcs from $\nVert[0]$ to $\nVert[k]$. Let $\nVertLvl[i]$ denote a set of vertices for which the distance from root vertex $\nVert[0]$ is $i$. Note that $\nVertLvl[0] = \{\nVert[0]\}$.

\subsection{Graph Generation}

The schedule graph is generated based on a set of tasks $\nEtTask$ and scheduling policy $\nPolicy$. The generation algorithm initializes with root vertex $\nVert[0]$ with $\nInt[0] = [0,0]$ and $\nVertLvl[0] = \{\nVert[0]\}$. It then keeps performing two alternating phases called \textit{expansion phase} and \textit{merge phase}. The expansion phase generates a new set of vertices $\nVertLvlExp[i+1]$ from $\nVertLvl[i]$. The merge phase then tries to merge some vertices from $\nVertLvlExp[i+1]$ with each other which results in $\nVertLvl[i+1]$.

The two alternating phases are performed until $\nVertLvl[n']$ is generated, where $n' = \sum_{i'=1}^{n}h_{i'}$ is the total number of jobs in all the tasks in $\nEtTask$. The algorithm may also end earlier if it finds a deadline miss. If a deadline miss is detected, a set of tasks $\nEtTask$ is not schedulable under policy $\nPolicy$. If the entire schedule graph is generated with no deadline misses found, then the set of tasks $\nEtTask$ is schedulable under policy $\nPolicy$. 

\subsubsection{Expansion Phase}

A path taken from the root vertex to vertex $\nVert[k]$ determines jobs that are finished in vertex $\nVert[k]$. Notice that all paths from the root vertex to vertex $\nVert[k]$ contain the same set of finished jobs. This set of finished jobs is denoted as $\nEdgePath[k]$ and is used to determine applicable jobs for vertex $\nVert[k]$.
The expansion phase evaluates each vertex in $\nVertLvl[i]$ individually by using its applicable jobs and its finish time interval $\nInt$. It determines at what time which jobs can be executed next. This in turn means generating new vertices and arcs (more details later in Section \ref{III.E.2}). The result is a set of new vertices $\nVertLvlExp[i+1]$, some of which may be merged in the merge phase.

\subsubsection{Merge Phase}

The merge phase takes the result of the expansion phase $\nVertLvlExp[i+1]$ and attempts to merge its vertices. Two vertices $\nVert[k]$ and $\nVert[k']$ can be merged if $\nEdgePath[k] = \nEdgePath[k'] \wedge \nInt[k] \cap \nInt[k'] \neq \emptyset$. If vertices $\nVert[k]$ and $\nVert[k']$ are merged, then all arcs pointing to $\nVert[k']$ are reoriented so that they point to $\nVert[k]$, time interval of $\nVert[k]$ is updated as $\nInt[k] \gets \nInt[k] \cup \nInt[k']$ and $\nVert[k']$ is removed from the graph. Note that the schedule graph is not a multigraph. %, which means that some arcs that are reoriented from $\nVert[k']$ to $\nVert[k]$ may be removed instead.

\subsubsection{Example}\label{exampleschedulegraph}

\begin{figure}[!hbt]
\centering
\begin{tikzpicture}
\pgfmathsetlengthmacro\MajorTickLength{
      \pgfkeysvalueof{/pgfplots/major tick length} * 0.5}
		\begin{axis}[
	        width=10cm,
	        height=4cm,
	        xmajorgrids={true},
			major grid style={densely dotted},
			ytick style={draw=none},
            axis lines = left,
            axis line style=-,
			enlarge x limits=0,
			enlarge y limits=0,
			tick align=outside,
			major tick length=\MajorTickLength,
			ymin=0,
			ymax=3,
			xmin=0,
			xmax=10,
			ytick={0.5,1.5,2.5},
			xtick={0,1,2,...,10},
			yticklabels={$\nEtTask[3]$,$\nEtTask[2]$,$\nEtTask[1]$},
			x tick label style={font=\scriptsize},
			y tick label style={font=\small}]
   \pozaditmave{0}{10}{2}
   \pozaditmave{0}{3}{1}
   \pozaditmave{5}{8}{1}
   \pozaditmave{3}{9}{0}
   \pozadisvetle{1}{3}{0}
   \deadline{10}{2}
   \deadline{10}{1}
   \deadline{5}{1}
   \deadline{10}{0}
   \jobjitter{1}{2}{2}
   \jobjitter{6}{7}{0}
   \job{0}{1}{2}{\nEtJob{1}{1}}
   \job{0}{1}{1}{\nEtJob{2}{1}}
   \job{5}{6}{1}{\nEtJob{2}{2}}
   \job{3}{6}{0}{\nEtJob{3}{1}}
\end{axis}
\end{tikzpicture}
    \caption{A loose chart of the illustrative instance given in Table~\ref{table:ex3_table}.}
    \label{fig:ex3_loose}
\end{figure}
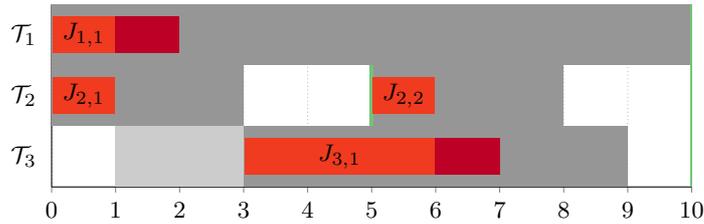

Consider the following illustrative problem instance consisting of 3 tasks $\nEtTask = (\nEtTask[1],\nEtTask[2],\nEtTask[3])$ with the parameters given in Table~\ref{table:ex3_table} and using the EDF scheduling policy. The priority values $\nPrEtTask$ are not specified as EDF does not make use of these values. The loose chart of this instance can be seen in \figureautorefname{}~\ref{fig:ex3_loose}.

\begin{table}
    \caption{Parameters of tasks used in the illustrative instance}
    \label{table:ex3_table}
    \centering
    \begin{tabular}{c c c c c c c} 
    \toprule
    & $\nRMin$ & $\nRMax$ & $\nCMin$ & $\nCMax$ & $\nDEtTask$ & $\nPeEtTask$ \\
    \midrule
    $\nEtTask[1]$ & 0 & 0 & 1 & 2 & 10 & 10 \\
    $\nEtTask[2]$ & 0 & 0 & 1 & 1 & 3 & 5 \\
    $\nEtTask[3]$ & 1 & 3 & 3 & 4 & 9 & 10 \\
    \bottomrule
    \end{tabular}
\end{table}

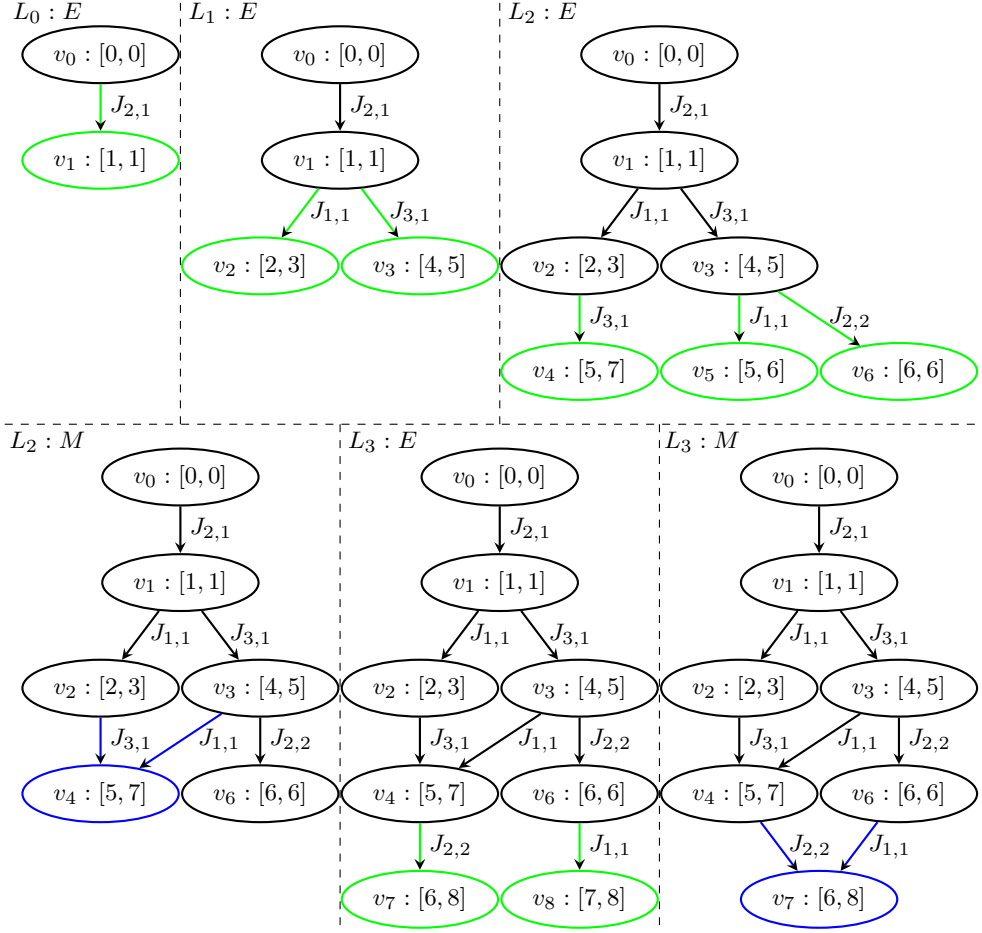
\begin{figure*}
    \centering
    \scriptsize
    \newcommand{\vertex}[6] { % id1, id2, eft, lft, csx, csy
 	\node[ellipse, draw, thick, minimum size=0.1cm] (v#1#2) at (#5,#6) {$v_{#1}: [#3,#4]$};
}
\newcommand{\greenvertex}[6] { % id1, id2, eft, lft, csx, csy
 	\node[ellipse, draw, thick, minimum size=0.1cm, draw=green] (v#1#2) at (#5,#6) {$v_{#1}: [#3,#4]$};
}
\newcommand{\bluevertex}[6] { % id1, id2, eft, lft, csx, csy
 	\node[ellipse, draw, thick, minimum size=0.1cm, draw=blue] (v#1#2) at (#5,#6) {$v_{#1}: [#3,#4]$};
}
%offset between levels
\newcommand{\ls}{2}

\begin{tikzpicture}[scale=0.7, draw=black, text=black]

\greenvertex{7}{5}{6}{8}{-1.5}{0}
\greenvertex{8}{5}{7}{8}{1.5}{0}
\bluevertex{7}{6}{6}{8}{6}{0}

\bluevertex{4}{4}{5}{7}{-7.5}{\ls}
\vertex{6}{4}{6}{6}{-4.5}{\ls}
\vertex{4}{5}{5}{7}{-1.5}{\ls}
\vertex{6}{5}{6}{6}{1.5}{\ls}
\vertex{4}{6}{5}{7}{4.5}{\ls}
\vertex{6}{6}{6}{6}{7.5}{\ls}

\draw[-stealth, thick, draw=green] (v45) -- node[anchor=west] {$\nEtJobS_{2,2}$} (v75);
\draw[-stealth, thick, draw=green] (v65) -- node[anchor=west] {$\nEtJobS_{1,1}$} (v85);
\draw[-stealth, thick, draw=blue] (v46) -- node[anchor=west] {$\nEtJobS_{2,2}$} (v76);
\draw[-stealth, thick, draw=blue] (v66) -- node[anchor=west] {$\nEtJobS_{1,1}$} (v76);

\vertex{2}{4}{2}{3}{-7.5}{\ls+\ls}
\vertex{3}{4}{4}{5}{-4.5}{\ls+\ls}
\vertex{2}{5}{2}{3}{-1.5}{\ls+\ls}
\vertex{3}{5}{4}{5}{1.5}{\ls+\ls}
\vertex{2}{6}{2}{3}{4.5}{\ls+\ls}
\vertex{3}{6}{4}{5}{7.5}{\ls+\ls}

\draw[-stealth, thick, draw=blue] (v24) -- node[anchor=west] {$\nEtJobS_{3,1}$} (v44);
\draw[-stealth, thick, draw=blue] (v34) -- node[anchor=west] {~$\nEtJobS_{1,1}$} (v44);
\draw[-stealth, thick] (v34) -- node[anchor=west] {$\nEtJobS_{2,2}$} (v64);
\draw[-stealth, thick] (v25) -- node[anchor=west] {$\nEtJobS_{3,1}$} (v45);
\draw[-stealth, thick] (v35) -- node[anchor=west] {~$\nEtJobS_{1,1}$} (v45);
\draw[-stealth, thick] (v35) -- node[anchor=west] {$\nEtJobS_{2,2}$} (v65);
\draw[-stealth, thick] (v26) -- node[anchor=west] {$\nEtJobS_{3,1}$} (v46);
\draw[-stealth, thick] (v36) -- node[anchor=west] {~$\nEtJobS_{1,1}$} (v46);
\draw[-stealth, thick] (v36) -- node[anchor=west] {$\nEtJobS_{2,2}$} (v66);

\vertex{1}{4}{1}{1}{-6}{\ls+\ls+\ls}
\vertex{1}{5}{1}{1}{0}{\ls+\ls+\ls}
\vertex{1}{6}{1}{1}{6}{\ls+\ls+\ls}

\draw[-stealth, thick] (v14) -- node[anchor=west] {$\nEtJobS_{1,1}$} (v24);
\draw[-stealth, thick] (v14) -- node[anchor=west] {$\nEtJobS_{3,1}$} (v34);
\draw[-stealth, thick] (v15) -- node[anchor=west] {$\nEtJobS_{1,1}$} (v25);
\draw[-stealth, thick] (v15) -- node[anchor=west] {$\nEtJobS_{3,1}$} (v35);
\draw[-stealth, thick] (v16) -- node[anchor=west] {$\nEtJobS_{1,1}$} (v26);
\draw[-stealth, thick] (v16) -- node[anchor=west] {$\nEtJobS_{3,1}$} (v36);

\vertex{0}{4}{0}{0}{-6}{\ls+\ls+\ls+\ls}
\vertex{0}{5}{0}{0}{0}{\ls+\ls+\ls+\ls}
\vertex{0}{6}{0}{0}{6}{\ls+\ls+\ls+\ls}

\draw[-stealth, thick] (v04) -- node[anchor=west] {$\nEtJobS_{2,1}$} (v14);
\draw[-stealth, thick] (v05) -- node[anchor=west] {$\nEtJobS_{2,1}$} (v15);
\draw[-stealth, thick] (v06) -- node[anchor=west] {$\nEtJobS_{2,1}$} (v16);

\greenvertex{4}{3}{5}{7}{1.5}{\ls+\ls+\ls+\ls+\ls}
\greenvertex{5}{3}{5}{6}{4.5}{\ls+\ls+\ls+\ls+\ls}
\greenvertex{6}{3}{6}{6}{7.5}{\ls+\ls+\ls+\ls+\ls}

\vertex{2}{3}{2}{3}{1.5}{\ls+\ls+\ls+\ls+\ls+\ls}
\vertex{3}{3}{4}{5}{4.5}{\ls+\ls+\ls+\ls+\ls+\ls}
\greenvertex{2}{2}{2}{3}{-4.5}{\ls+\ls+\ls+\ls+\ls+\ls}
\greenvertex{3}{2}{4}{5}{-1.5}{\ls+\ls+\ls+\ls+\ls+\ls}

\draw[-stealth, thick, draw=green] (v23) -- node[anchor=west] {$\nEtJobS_{3,1}$} (v43);
\draw[-stealth, thick, draw=green] (v33) -- node[anchor=west] {$\nEtJobS_{1,1}$} (v53);
\draw[-stealth, thick, draw=green] (v33) -- node[anchor=west] {$\nEtJobS_{2,2}$} (v63);

\greenvertex{1}{1}{1}{1}{-7.5}{\ls+\ls+\ls+\ls+\ls+\ls+\ls}
\vertex{1}{2}{1}{1}{-3}{\ls+\ls+\ls+\ls+\ls+\ls+\ls}
\vertex{1}{3}{1}{1}{3}{\ls+\ls+\ls+\ls+\ls+\ls+\ls}

\draw[-stealth, thick, draw=green] (v12) -- node[anchor=west] {$\nEtJobS_{1,1}$} (v22);
\draw[-stealth, thick, draw=green] (v12) -- node[anchor=west] {$\nEtJobS_{3,1}$} (v32);
\draw[-stealth, thick] (v13) -- node[anchor=west] {$\nEtJobS_{1,1}$} (v23);
\draw[-stealth, thick] (v13) -- node[anchor=west] {$\nEtJobS_{3,1}$} (v33);

\vertex{0}{1}{0}{0}{-7.5}{\ls+\ls+\ls+\ls+\ls+\ls+\ls+\ls}
\vertex{0}{2}{0}{0}{-3}{\ls+\ls+\ls+\ls+\ls+\ls+\ls+\ls}
\vertex{0}{3}{0}{0}{3}{\ls+\ls+\ls+\ls+\ls+\ls+\ls+\ls}

\draw[-stealth, thick, draw=green] (v01) -- node[anchor=west] {$\nEtJobS_{2,1}$} (v11);
\draw[-stealth, thick] (v02) -- node[anchor=west] {$\nEtJobS_{2,1}$} (v12);
\draw[-stealth, thick] (v03) -- node[anchor=west] {$\nEtJobS_{2,1}$} (v13);

\draw[dashed] (-9.3,\ls+\ls+\ls+\ls+\ls/2) -- (9,\ls+\ls+\ls+\ls+\ls/2);
\draw[dashed] (-3,\ls+\ls+\ls+\ls+\ls/2) -- (-3,-\ls/3);
\draw[dashed] (3,\ls+\ls+\ls+\ls+\ls/2) -- (3,-\ls/3);
\draw[dashed] (-6,\ls+\ls+\ls+\ls+\ls/2) -- (-6,\ls+\ls+\ls+\ls+\ls+\ls+\ls+\ls+\ls/2);
\draw[dashed] (0,\ls+\ls+\ls+\ls+\ls/2) -- (0,\ls+\ls+\ls+\ls+\ls+\ls+\ls+\ls+\ls/2);

\node at (-8.5,\ls+\ls+\ls+\ls+\ls+\ls+\ls+\ls+\ls/2.5) {$L_{0}: E$};
\node at (-5.2,\ls+\ls+\ls+\ls+\ls+\ls+\ls+\ls+\ls/2.5) {$L_{1}: E$};
\node at (0.8,\ls+\ls+\ls+\ls+\ls+\ls+\ls+\ls+\ls/2.5) {$L_{2}: E$};
\node at (-8.5,\ls+\ls+\ls+\ls+\ls/3) {$L_{2}: M$};
\node at (-2.2,\ls+\ls+\ls+\ls+\ls/3) {$L_{3}: E$};
\node at (3.8,\ls+\ls+\ls+\ls+\ls/3) {$L_{3}: M$};

\end{tikzpicture}

    \caption{The process of building a schedule graph using EDF policy. Each vertex $\nVert[k]$ contains its finish time interval $\nInt[k]$. Each arc is labeled by the executed job. The stages are annotated with level and type of phase (either E for expansion or M for merge). The merge phases on levels 0 and 1 do not change the graph and are therefore omitted.}
    \label{fig:ex3_graph}
\end{figure*}

The entire schedule graph generation process is visualized in \figureautorefname{}~\ref{fig:ex3_graph}. The algorithm starts at level 0 with root vertex $\nVert[0]$, $\nInt[0] = [0,0]$ and no jobs finished. At this point, there are two certainly released jobs $\nEtJob{1}{1}$ and $\nEtJob{2}{1}$. The expansion phase concludes that the EDF policy would pick $\nEtJob{2}{1}$ at time $0$ as it has an earlier deadline than $\nEtJob{1}{1}$. Job $\nEtJob{3}{1}$ is not considered as it would not be released at time $0$ for any execution scenario. The expansion results in creating a new vertex $\nVert[1]$ with $\nInt[1] = [\nEft[0] + \nCMinEtJob{2}{1}, \nLft[0] + \nCMaxEtJob{2}{1}] = [1,1]$. Additionally, new arc pointing from $\nVert[0]$ to $\nVert[1]$ is created and labeled with the picked job $\nEtJob{2}{1}$. This concludes the expansion phase on level 0. The merge phase is performed next but it does not change the graph in any way as the result of the expansion phase is only one vertex.

The algorithm continues with the expansion phase on level 1 with $\nVert[1]$. Here, two scenarios may occur. In the first scenario, job $\nEtJob{3}{1}$ has just released at time $1$. If that is the case, $\nEtJob{3}{1}$ would be picked by EDF policy as it has an earlier deadline than $\nEtJob{1}{1}$. In the second scenario, job $\nEtJob{3}{1}$ does not release at time $1$, and $\nEtJob{1}{1}$ is picked instead as there are no other released jobs. These two scenarios result in vertices $\nVert[2]$ and $\nVert[3]$ with $\nInt[2] = [\nEft[1] + \nCMinEtJob{1}{1}, \nLft[1] + \nCMaxEtJob{1}{1}] = [2,3]$ and $\nInt[3] = [\nEft[1] + \nCMinEtJob{3}{1}, \nLft[1] + \nCMaxEtJob{3}{1}] = [4,5]$. This is the result of the expansion phase. The merge phase does not merge vertices $\nVert[2]$ and $\nVert[3]$ as $\nEdgePath[2] \neq \nEdgePath[3]$.

At level 2, the expansion phase individually evaluates $\nVert[2]$ and $\nVert[3]$. When expanding a vertex, which has the earliest finish time strictly smaller than the latest finish time, we need to check all time points in the interval $\nInt$. In the case of $\nVert[2]$, the only jobs left are $\nEtJob{3}{1}$ and $\nEtJob{2}{2}$, but $\nEtJob{2}{2}$ will not release until time $5$ and is therefore ignored. At time $2$, there are two scenarios. If $\nEtJob{3}{1}$ releases at time $2$, then it would be picked by EDF policy. If $\nEtJob{3}{1}$ releases at time $3$, then the processor would stay idle until time $3$ (the online scheduler would be invoked at time $3$). At time $3$, the job $\nEtJob{3}{1}$ is certainly released and would be picked by EDF policy. Hence, EDF will always pick job $\nEtJob{3}{1}$ in time interval $[2,3]$. The earliest time it would start its execution is $2$ and the latest time is $3$. This results in a new vertex $\nVert[4]$ with $\nInt[4] = [\nEft[2] + \nCMinEtJob{3}{1}, \nLft[2] + \nCMaxEtJob{3}{1}] = [5,7]$.

The expansion phase proceeds to expand $\nVert[3]$. Vertex $\nVert[3]$ has finish time interval $\nInt[3] = [4,5]$. At time $4$, job $\nEtJob{1}{1}$ would be picked by EDF because $\nEtJob{2}{2}$ is not yet released and there are no other unfinished jobs. However, at time $5$, job $\nEtJob{2}{2}$ is certainly released and because it has a lower deadline than $\nEtJob{1}{1}$, $\nEtJob{2}{2}$ would be picked by EDF. This results in two new vertices $\nVert[5]$ and $\nVert[6]$, where $\nInt[5] = [4 + \nCMinEtJob{1}{1}, 4 + \nCMaxEtJob{1}{1}] = [5,6]$ and $\nInt[6] = [5 + \nCMinEtJob{2}{2}, 5 + \nCMaxEtJob{2}{2}] = [6,6]$. Notice that since different jobs would execute at different times, the new finish time intervals of $\nVert[5]$ and $\nVert[6]$ are computed from the earliest and latest time they would be picked by EDF.

After the expansion phase on level 2 is done, the merge phase takes over. Since $\nEdgePath[4] = \nEdgePath[5]$ and $\nInt[4] \cap \nInt[5] \neq \emptyset$, vertices $\nVert[4]$ and $\nVert[5]$ are merged. This is done by merging the intervals, arcs pointing to $\nVert[5]$ are redirected to $\nVert[4]$, and $\nVert[5]$ is removed. The resulting interval is $\nInt[4] = [5,7] \cup [5,6] = [5,7]$.

On level 3, the final iteration of the expansion and merge phase is performed. The expansion phase individually evaluates $\nVert[4]$ and $\nVert[6]$. For $\nVert[4]$, the last remaining job is $\nEtJob{2}{2}$, which releases at time $5$, and for $\nVert[6]$, the last remaining job is $\nEtJob{1}{1}$, which released at time $0$. The expansion phase creates vertex $\nVert[7]$ as expansion of $\nVert[4]$ and $\nVert[8]$ as expansion of $\nVert[6]$. The resulting finish time intervals are $\nInt[7] = [\nEft[4] + \nCMinEtJob{2}{2}, \nLft[4] + \nCMaxEtJob{2}{2}] = [6,8]$ and $\nInt[8] = [\nEft[6] + \nCMinEtJob{1}{1}, \nLft[6] + \nCMaxEtJob{1}{1}] = [7,8]$.

Since $\nEdgePath[7] = \nEdgePath[8]$ and $\nInt[7] \cap \nInt[8] \neq \emptyset$, vertices $\nVert[7]$ and $\nVert[8]$ are merged. This results in $\nVert[7]$ being the only vertex on level 3 with $\nInt[7] = [6,8]$.

There was no deadline miss in this example, therefore, the given set of tasks is schedulable under the EDF scheduling policy. A deadline miss could be easily detected during the expansion phase. When creating new vertex $\nVert[k]$ with job $\nEtJob{i}{j}$, there is a deadline miss if $\nLft[k] > \nDEtJob{i}{j}$.

\subsection{Detailed Description of Schedule Graph Generation}

The structure of the algorithm so far is the same as that of SGA-SE \cite{SANS}. The main novelty in our algorithm SGA-ME lies in the schedule graph generation rules for expanding vertices.
To ease the description, we will use the dot notation hereinafter. For instance, the deadline of job $\nEtJobS$ will be denoted as $\nEtJobS.\nD$.

Let us first define the parameters of each vertex and arc. Every vertex $\nVert$ has earliest finish time $\nVertEft$, latest finish time $\nVertLft$, incoming arcs $\nVertIn$ and outgoing arcs $\nVertOut$. Every arc $\nEdge$ has a job label $\nEdge.\nEtJobS$, source vertex $\nEdge.s$ and destination vertex $\nEdge.d$.

As mentioned before, we can determine the finished jobs of vertex $\nVert$ from its position in the graph. This is done by taking any directed path from root vertex $\nVert[0]$ to vertex $\nVert$ and then transforming the set of arcs on the path into a set of jobs by taking the labels of the arcs. This set of finished jobs is then used to determine the applicable jobs. The set of applicable jobs of vertex $\nVert$ is denoted as $\nVertAppJobs$.

\subsubsection{Job Eligibility for Work-Conserving Policies}

In this subsection, we deal with the rules for work-conserving policies.
Recall that the scheduling policy is a function $\nPolicy(t,\nAppJobs)$. We say that job $\nEtJob{i}{j}$ has higher \hbox{$\nPolicy$-priority} than $\nEtJob{x}{y}$ if $\nPolicy$($\myInf$, \{$\nEtJob{i}{j}$, $\nEtJob{x}{y}$\}) returns job $\nEtJob{i}{j}$. That is, $\nEtJob{i}{j}$ would be picked by the scheduling policy given that both $\nEtJob{i}{j}$ and $\nEtJob{x}{y}$ are certainly released. If one of the jobs is null, then the scheduling policy picks the one that is not null. %Behavior for both jobs being null is undefined. 
The policy function $\nPolicy$($\myInf$, \{$\nEtJob{i}{j}$, $\nEtJob{x}{y}$\}) can be replaced with any specific policy, e.g., FP-EDF($\myInf$, \{$\nEtJob{i}{j}$, $\nEtJob{x}{y}$\}).

For vertex $\nVert$, set of applicable jobs $\nVertAppJobs$ and time $t$, job $\nCeJob[t] \in \nVertAppJobs$ is \textit{certainly-eligible} at time $t$ if $\nCeJob[t]$ is certainly released at time $t$ and there is no other certainly released applicable job at time $t$ with higher $\nPolicy$-priority than $\nCeJob[t]$.
Formally, job $\nCeJob[t] \in \nVertAppJobs$ is certainly-eligible at time $t$ iff
\begin{equation}\label{def:ce}
\nCeJob[t].r^{max} \le t \wedge \nexists \nEtJob{i}{j} \in \nVertAppJobs \; st. \; \nEtJob{i}{j} \not= \nCeJob[t] \; \wedge \nEtJobRMax{i}{j} \leq t \wedge \nEtJob{i}{j} =  \nPolicy(\myInf, \{\nEtJob{i}{j}, \nCeJob[t]\})
\end{equation}
Note that there may be at most one certainly-eligible job at time $t$. If there is no certainly-eligible job at time $t$, we say that $\nCeJob[t]$ does not exist.

Similarly, job $\nPeJob[t] \in \nVertAppJobs$ is \textit{possibly-eligible} at time $t$ if it is possibly released at time $t$ and $\nCeJob[t]$ (if exists) has lower $\nPolicy$-priority than $\nPeJob[t]$. Formally, we define the set of possibly-eligible jobs at time $t$ as
\begin{equation}\label{def:pe}
\{ \nEtJob{i}{j} \mid \nEtJob{i}{j} \in \nVertAppJobs \; \wedge \; \nEtJobRMin{i}{j} \leq t < \nEtJobRMax{i}{j} \; \wedge \nEtJob{i}{j} = \nPolicy(\myInf, \{\nEtJob{i}{j}, \nCeJob[t]\}) \}
\end{equation}

\begin{table}[htp]
\caption{An example with 7 jobs (each marked with $\times$) of different priorities and release status: a certainly-eligible job is marked with $\boxtimes$ and possibly-eligible jobs are marked with $\otimes$}
\label{tab:cepeex}
\scriptsize
\centering
\begin{tabular}{cccc}

    \toprule
$\nPolicy$-priority & $t < \nEtJobRMin{i}{j}$ & $\nEtJobRMin{i}{j} \le t < \nEtJobRMax{i}{j}$ & $\nEtJobRMax{i}{j} \le t$ \\ 
 of $\nEtJob{i}{j}$ &  (not released) & (possibly released) & (certainly released) \\

\midrule 

0 & & $\otimes$ &  \\
1 & $\times$ & &  \\
2 & & $\otimes$ &  \\
3 & & & $\boxtimes$  \\
4 & & & $\times$  \\
5 & $\times$ & &  \\
6 & & $\times$ &  \\
    \bottomrule

\end{tabular}
\end{table}

To understand the definitions better, consider the illustration in Table~\ref{tab:cepeex} with 7 applicable jobs ordered top down by the $\nPolicy$-priority (0 for the highest $\nPolicy$-priority, 6 for the lowest). Recall that the scheduling policy is assumed to be deterministic, so there is a linear ordering on the jobs concerning the $\nPolicy$-priority. Each line then represents one job. There is a symbol $\times$ for each job according to whether the job is certainly released (3rd column), possibly released (2nd column), or neither (1st column). The certainly-eligible job, marked with the symbol $\boxtimes$, is the one with the highest priority from the column of certainly released jobs. The possibly-eligible jobs, marked with the symbol $\otimes$, are those jobs from the column of possibly released jobs that have higher priority than the certainly-eligible job.

The certainly-eligible job $\nCeJob[t]$ is the job that would be picked by the policy function at time $t$ in such a scenario when all possibly-eligible jobs are released at time $t+1$ or later, because in this scenario, all other jobs, which are neither certainly-eligible nor possibly-eligible at time $t$, either have lower $\nPolicy$-priority than $\nCeJob[t]$ or cannot be released at time $t$.

If one or more of the possibly-eligible jobs at time $t$ release at time $t$ or earlier, then one of these jobs is picked by the policy function at time $t$ because the possibly-eligible job has higher $\nPolicy$-priority than the certainly-eligible job $\nCeJob[t]$. However, we do not know in advance which possibly-eligible jobs will be released at time $t$ or earlier. Any of the possibly-eligible jobs may release at time $t$ or earlier while all other possibly-eligible jobs release at time $t+1$ or later. This means that each of the possibly-eligible jobs may be picked by the policy function in some scenarios.

Now, we explain how we create the exploration interval that is used for finding eligible jobs and thus for creating the next nodes. Given a set of applicable jobs $\nVertAppJobs$ and a work-conserving scheduling policy, jobs that are certainly-eligible or possibly-eligible at time $t \in [\nVertEft,\nVertLft]$ may begin execution at time $t$ as the next job in some execution scenario. Since there may be no certainly released job in the interval $[\nVertEft,\nVertLft]$, we extend the interval to a time point when a certainly-eligible job exists. Hence, the exploration interval is $[\nVertEft, \nLftExt]$, where $\nLftExt = min\{t \mid t \geq \nVertLft \wedge \nCeJob[t] \; exists\}$.

\subsubsection{Job Eligibility for Non-Work-Conserving Policies}

In this subsection, we extend the eligibility rules for non-work-conserving policies (e.g., P-FP-EDF).
What the considered non-work-conserving policies have in common is that they all entail some critical job $\nCritJob$ and critical time $\nCritTime$, where $\nCritTime$ is the latest time when $\nCritJob$ must be started to avoid missing its deadline. Hence, no job, except for $\nCritJob$, can start its execution if it may finish after $\nCritTime$. How to obtain $\nCritTime$ and $\nCritJob$ is described in \ref{sec:policies}.

While the original SGA-SE \cite{SANS} aims to work with the special class of non-work-conserving policies that augment the job priority ordering with an idle-time insertion policy (IIP) \cite{SANS}, which the authors adjust to job-set-dependent IIPs, our algorithm SGA-ME allows working with any non-work-conserving policy from which $\nCritTime$ and $\nCritJob$ can be inferred. 

We redefine certainly and possibly-eligible jobs based on vertex $\nVert$, set of applicable jobs $\nVertAppJobs$, critical time $\nCritTime$, critical job $\nCritJob$ and time $t$. %We defined them based on vertex $\nVert$ and time $t$.
Given vertex $\nVert$ and time $t$, job $\nEtJob{i}{j} \in \nVertAppJobs$ \textit{violates} $\nVertCritTime$ if $t + \nEtJobCMax{i}{j} > \nVertCritTime \wedge \nEtJob{i}{j} \neq \nVertCritJob$. That is, $\nEtJob{i}{j}$ violates $\nVertCritTime$ if $\nEtJob{i}{j}$ is not the critical job and it may finish after $\nVertCritTime$ when started at time $t$.

Now, \textit{certainly-eligible} job is the one with the highest $\nPolicy$-priority from the set of certainly released jobs that do not violate $\nVertCritTime$.
Formally, given vertex $\nVert$, job $\nCeJob[t] \in \nVertAppJobs$ is certainly-eligible at time $t$ iff
%\small
\begin{multline}\label{def:newce}
\nCeJob[t].r^{max} \le t ~\wedge~ \underline{(t + \nCeJob[t].C^{max} \leq \nVertCritTime \vee \nCeJob[t] = \nVertCritJob)} ~\wedge~ \nexists \nEtJob{i}{j} \in \nVertAppJobs \; \text{st.} \\
\nEtJob{i}{j} \not= \nCeJob[t] ~\wedge~ \nEtJobRMax{i}{j} \leq t ~\wedge~ \nEtJob{i}{j} = \nPolicy(\myInf, \{\nEtJob{i}{j}, \nCeJob[t]\}) ~\wedge~ \underline{(t + \nEtJobCMax{i}{j} \leq \nVertCritTime \vee \nEtJob{i}{j} = \nVertCritJob)}
\end{multline}
%\normalsize

The underlined parts emphasize what is added in Equation~(\ref{def:newce}) compared to Equation~(\ref{def:ce}). Similarly, job $\nPeJob[t]$ is \textit{possibly-eligible} at time $t$ if it is possibly released at time $t$, does not violate $\nVertCritTime$, and $\nCeJob[t]$ (if exists) has lower $\nPolicy$-priority than $\nPeJob[t]$. Formally, we define the set of possibly-eligible jobs at time $t$ as 
%\small
\begin{multline}\label{def:newpe}
\{ \nEtJob{i}{j} \mid \nEtJob{i}{j} \in \nVertAppJobs \; \wedge \; \nEtJobRMin{i}{j} \leq t < \nEtJobRMax{i}{j} \; \wedge\\
\nEtJob{i}{j} = \nPolicy(\myInf, \{\nEtJob{i}{j}, \nCeJob[t]\}) ~\wedge~ \underline{(t + \nEtJobCMax{i}{j} \leq \nVertCritTime \vee \nEtJob{i}{j} = \nVertCritJob)} \}
\end{multline} 
%\normalsize

Again, the underlined part emphasizes what is added in Equation~(\ref{def:newpe}) compared to Equation~(\ref{def:pe}). Just as with work-conserving policies, jobs that are certainly-eligible or possibly-eligible at time $t \in \nVertInt$ may begin execution at time $t$ in some execution scenarios.

\subsubsection{Expansion Phase}\label{III.E.2}

The goal of the expansion phase is to find times on the exploration interval when each job is either possibly or certainly-eligible and expand vertices accordingly. The pseudocode of the expansion phase can be seen in Algorithm~\ref{exp_phase_edffp}. 

%\begin{fullwidth}[width=\linewidth+2cm,leftmargin=-1cm,rightmargin=-1cm]
\begin{algorithm}
\footnotesize
    \caption{Expansion Phase}\label{exp_phase_edffp}
    % \hspace*{\algorithmicindent} \textbf{Input:} set of vertices $\nVertLvl[i]$ \\
    % \hspace*{\algorithmicindent} \textbf{Output:} set of vertices $\nVertLvlExp[i+1]$ \\
    \begin{algorithmic}[1]
    \Function{EXPANSION\_PHASE}{$\nVertLvl[i]$}
    \State \textbf{$\nVertLvlExp[i+1] \gets$} $\emptyset$
    \For {\textbf{each} vertex $v \in \nVertLvl[i]$}
    \State \textbf{$\nVertLvlN[new] \gets$} NEXT\_NODES($\nVert$)
    \State $\nVertLvlExp[i+1] \gets \nVertLvlExp[i+1] \cup \nVertLvlN[new]$
    \EndFor
    \State \textbf{return} $\nVertLvlExp[i+1]$
    \EndFunction
    
    \Function{NEXT\_NODES}{$\nVert$}
    \If {$\nVertAppJobs = \emptyset$}
    \State \textbf{return} $\emptyset$
    \EndIf
    \State \textbf{$\nVertLvlN[new] \gets$} $\emptyset$
    \State \textbf{$\nLftExt \gets min\{t \mid t \geq \nVertLft \wedge \nCeJob[t] \; exists\}$}
    
    \For {\textbf{each} $\nEtJobS \in \nAppJobs$}
        \State \textbf{$t^{el} \gets$} integer time points $t \in [\nVertEft,\nLftExt]$ when $\nEtJobS$ is possibly or certainly-eligible\label{exp_phase_edffp_time}
        \State \textbf{$t^r \gets$} times $t^{el}$ converted into ranges\label{exp_phase_edffp_range_conversion}
        \For{\textbf{each} range $[est,lst] \in t^r$}
        \State \textbf{$\nVert[k] \gets$} EXPAND($v,\nEtJobS,est,lst$)
        \State $\nVertLvlN[new] \gets \nVertLvlN[new] \cup \{\nVert[k]\}$
        \EndFor
    \EndFor
    \State \textbf{return} $\nVertLvlN[new]$
    
    \EndFunction
    
    \Function{EXPAND}{$\nVert, \nEtJobS, est, lst$}
        \State \textbf{$\nVert[k] \gets$} new vertex with $\nVertEft[k] = est + \nEtJobS.\nCMin$ and $\nVertLft[k] = lst + \nEtJobS.\nCMax$
        \State \textbf{$\nEdge \gets$} new arc with $\nEdge.\nEtJobS = \nEtJobS$, $\nEdge.s = \nVert$ and $\nEdge.d = \nVert[k]$
        \State $\nVertIn[k] \gets \{\nEdge\}$
        \State $\nVertOut \gets \nVertOut \cup \{\nEdge\}$
        \State \textbf{return} $\nVert[k]$
    \EndFunction
    
    \end{algorithmic}
\end{algorithm}
%\end{fullwidth}

Notice the conversion of integers into ranges on line \ref{exp_phase_edffp_range_conversion}. This is done such that, for illustration, the set of integers \{2,3,4,5,7,8,10,14,15,16\} is converted into the set of ranges \{[2,5], [7,8], [10,10], [14,16]\}.
Due to the properties of work-conserving policies, there will always be only one range in variable $t^r$. In addition, the left boundary of this range $est$ will always be equal to $max(\nVertEft,\nEtJobRMin{i}{j})$. These two observations do not hold for non-work-conserving policies as will be seen further in Section~\ref{app_A_sans_error}.

A more detailed description of function NEXT\_NODES for work-conserving policies is provided in \ref{sec:nextnodesimpl}.

\subsubsection{Merge Phase}

Unlike the expansion phase, the merge phase is the same for both work-conserving and non-work-conserving policies. A pseudocode of the merge phase can be seen in Algorithm \ref{merge_phase}.

\begin{algorithm}
\footnotesize
    \caption{Merge Phase}\label{merge_phase}
    % \hspace*{\algorithmicindent} \textbf{Input:} a set of vertices $\nVertLvlExp[i]$ \\
    % \hspace*{\algorithmicindent} \textbf{Output:} none, the changes are done locally \\
    \begin{algorithmic}[1]
    
    \Function{MERGE\_PHASE}{$\nVertLvlExp[i]$}
    
    \While{$\exists \nVert[k],\nVert[x] \in \nVertLvlExp[i]$ st. $\nEdgePath[k] = \nEdgePath[x] \wedge \nVertInt[k] \cap \nVertInt[x] \neq \emptyset$}\label{ln:mergecondition}
    \State \textbf{$\nVertEft[k] \gets$} min($\nVertEft[k]$, $\nVertEft[x]$)
    \State \textbf{$\nVertLft[k] \gets$} max($\nVertLft[k]$, $\nVertLft[x]$)
    \For {\textbf{each} arc $\nEdge \in \nVertIn[x]$}
    \If{$ \nexists \nEdge ' \in \nVertIn[k]$ st. $\nEdge.s = \nEdge '.s$}\label{multigraphcondition}
    \State $\nEdge.d = \nVert[k]$
    \State $\nVertIn[k] = \nVertIn[k] \cup \{\nEdge\}$
    \EndIf
    \EndFor
    \State remove $\nVert[x]$ and all arcs \{$\nEdge \mid \nEdge.d = \nVert[x]$\}
    \EndWhile
    \EndFunction
    
    \end{algorithmic}
\end{algorithm}

The merge phase keeps merging vertices until there are no two vertices that satisfy the conditions to be merged (line~\ref{ln:mergecondition}). When merging two vertices $\nVert[k]$ and $\nVert[x]$, vertex $\nVert[k]$ has its finish time range changed to $\nVertInt[k] \cup \nVertInt[x]$ and all arcs directed to $\nVert[x]$ are redirected to $\nVert[k]$. The condition on line \ref{multigraphcondition} serves to avoid creating a multigraph.

\subsubsection{Complete Schedule Graph Generation}

Schedule graph generation alternates between the described expansion and merge phases. It starts with root node $\nVert[0]$, $\nVertEft[0] = \nVertLft[0] = 0$ and $\nVertIn[0] = \nVertOut[0] = \emptyset$. The pseudocode of the graph generation algorithm can be seen in Algorithm \ref{graph_gen_edffp}.

\begin{algorithm}
\footnotesize
    \caption{Schedule Graph Generation}\label{graph_gen_edffp}
    % \hspace*{\algorithmicindent} \textbf{Input:} a set of tasks $\nEtTask$, policy function $\nPolicy$ \\
    % \hspace*{\algorithmicindent} \textbf{Output:} if set of tasks $\nEtTask$ is schedulable under a given policy \\
    \begin{algorithmic}[1]
    
    \Function{GRAPH\_GENERATION}{$\nEtTask$}
    \State \textbf{$\nVert[0] \gets$} new vertex with $\nVertEft[0] = \nVertLft[0] = 0$ and $\nVertIn[0] = \nVertOut[0] = \emptyset$
    \State \textbf{$\nVertLvl[0] \gets \{\nVert[0]\}$}
    \For {$i = 1$ \textbf{to} $\sum_{i'=1}^{n}h_{i'}$} \Comment{$\sum_{i'=1}^{n}h_{i'}$ is the total number of jobs in all the tasks in $\nEtTask$}
    \State \textbf{$\nVertLvlExp[i] \gets$} EXPANSION\_PHASE($\nVertLvl[i-1]$)
    \For {\textbf{each} $\nVert \in \nVertLvlExp[i]$} \label{graph_gen_edffp_dmd_start}
    \State \textbf{$\nEtJobS \gets$} label of the only arc in $\nVert.in$
    \If{$\nVert.\nLft > \nEtJobS.\nD$}
    \State \textbf{return} $false$ \Comment{Deadline miss detected}  \label{graph_gen_edffp_dmd_end}
    \EndIf
    \EndFor
    \State \textbf{$\nVertLvl[i] \gets$} MERGE\_PHASE($\nVertLvlExp[i]$)
    \EndFor
    \State \textbf{return} $true$ \Comment{Generation completed with no deadline misses}
    \EndFunction
    
    \end{algorithmic}
\end{algorithm}

The graph is generated and changed only in the expansion and merge phases. The function GRAPH\_GENERATION combines the two phases while detecting deadline misses.

The part of the code that detects deadlines (lines \ref{graph_gen_edffp_dmd_start}-\ref{graph_gen_edffp_dmd_end}) can be performed after the merge phase on a potentially smaller number of vertices. However, in the case of a deadline miss, the last merge phase would be redundant.
Note that the merge phase does not need to be done in the very last iteration of the algorithm. %Instead, the deadline miss detection would be executed directly on the expanded vertices $\nVertLvlExp[i]$ and the merge phase would be omitted as its result is not used.

\subsection{Difference of SGA-SE and SGA-ME} \label{app_A_sans_error}

In this section, we analyze the difference of the behavior of SGA-SE and SGA-ME on an example instance to understand the modified concept of non-work-conserving policies.

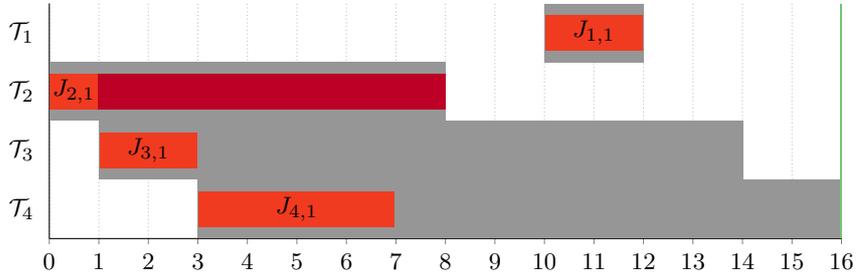
\begin{figure}[!hbt]
\centering
\begin{tikzpicture}
\pgfmathsetlengthmacro\MajorTickLength{
      \pgfkeysvalueof{/pgfplots/major tick length} * 0.5}
		\begin{axis}[
	        width=12cm,
	        height=4.7cm,
	        xmajorgrids={true},
			major grid style={densely dotted},
			ytick style={draw=none},
            axis lines = left,
            axis line style=-,
			enlarge x limits=0,
			enlarge y limits=0,
			tick align=outside,
			major tick length=\MajorTickLength,
			ymin=0,
			ymax=4,
			xmin=0,
			xmax=16,
			ytick={0.5,1.5,2.5,3.5},
			xtick={0,1,2,...,16},
			yticklabels={$\nEtTask[4]$,$\nEtTask[3]$,$\nEtTask[2]$,$\nEtTask[1]$},
			x tick label style={font=\scriptsize},
			y tick label style={font=\small}]
   \pozaditmave{10}{12}{3}
   \pozaditmave{0}{8}{2}
   \pozaditmave{1}{14}{1}
   \pozaditmave{3}{16}{0}
   \deadline{16}{3}
   \deadline{16}{2}
   \deadline{16}{1}
   \deadline{16}{0}
   \jobjitter{1}{8}{2}
   \job{10}{12}{3}{\nEtJob{1}{1}}
   \job{0}{1}{2}{\nEtJob{2}{1}}
   \job{1}{3}{1}{\nEtJob{3}{1}}
   \job{3}{7}{0}{\nEtJob{4}{1}}
\end{axis}
\end{tikzpicture}
    \caption{A loose chart of the example instance given in Table~\ref{table:sans_error_table}.}
    \label{fig:sans_error_gantt}
\end{figure}

Consider the following problem instance consisting of 4 tasks $\nEtTask = (\nEtTask[1],\nEtTask[2],\nEtTask[3],\nEtTask[4])$ whose parameters are listed in Table \ref{table:sans_error_table}. The release dates and computation times are deterministic except for different $\nCMin$ and $\nCMax$ of task $\nEtTask[2]$. The instance is visualized by the loose chart depicted in \figureautorefname{}~\ref{fig:sans_error_gantt}.

\setlength\extrarowheight{2pt}
\begin{table}
    \caption{The example instance parameters}
    \label{table:sans_error_table}
    \centering
    \begin{tabular}{c c c c c c c c}
    \toprule
    & $\nRMin$ & $\nRMax$ & $\nCMin$ & $\nCMax$ & $\nDEtTask$ & $\nPeEtTask$ & $\nPr$ \\
    \midrule
    $\nEtTask[1]$ & 10 & 10 & 2 & 2 & 12 & 16 & 0 \\
    $\nEtTask[2]$ & 0 & 0 & 1 & 8 & 8 & 16 & 1 \\
    $\nEtTask[3]$ & 1 & 1 & 2 & 2 & 14 & 16 & 3 \\
    $\nEtTask[4]$ & 3 & 3 & 4 & 4 & 16 & 16 & 2 \\
    \bottomrule
    \end{tabular}
\end{table}

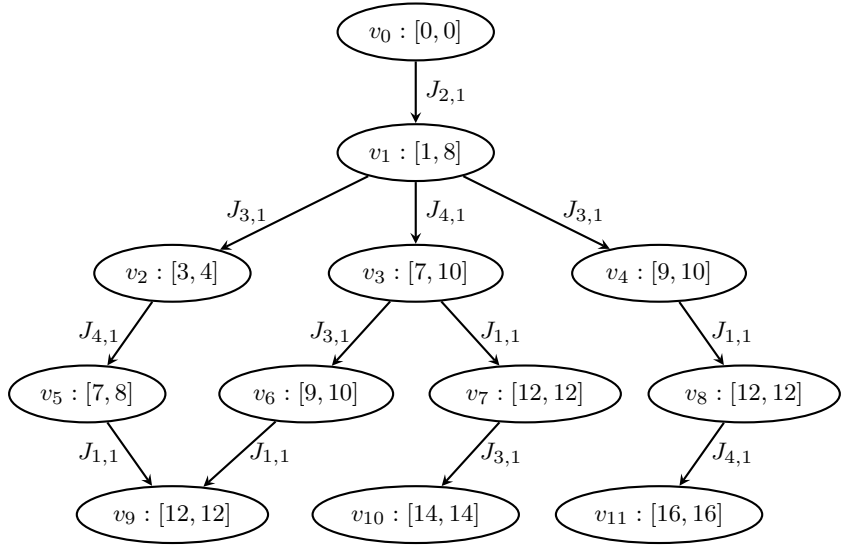
\begin{figure}[!hbt]
    % \includesvg[width=0.48\textwidth]{img/SANS_error/graph_correct.svg}
    \centering  
    \scriptsize
\newcommand{\vertex}[5] { % id, eft, lft, csx, csy
 	\node[ellipse, draw, thick, minimum size=0.1cm] (v#1) at (#4,#5) {$v_{#1}: [#2,#3]$};
}
\newcommand{\ls}{2}
\begin{tikzpicture}[scale=0.8, draw=black, text=black]
\vertex{0}{0}{0}{0}{0}
\vertex{1}{1}{8}{0}{-\ls}
\draw[-stealth, thick] (v0) -- node[anchor=west] {$\nEtJobS_{2,1}$} (v1);
\vertex{2}{3}{4}{-4}{-\ls-\ls}
\vertex{3}{7}{10}{0}{-\ls-\ls}
\vertex{4}{9}{10}{4}{-\ls-\ls}
\draw[-stealth, thick] (v1) -- node[anchor=east] {$\nEtJobS_{3,1}$~~~} (v2);
\draw[-stealth, thick] (v1) -- node[anchor=west] {$\nEtJobS_{4,1}$} (v3);
\draw[-stealth, thick] (v1) -- node[anchor=west] {~~$\nEtJobS_{3,1}$} (v4);
\vertex{5}{7}{8}{-5.4}{-\ls-\ls-\ls}
\vertex{6}{9}{10}{-1.8}{-\ls-\ls-\ls}
\vertex{7}{12}{12}{1.8}{-\ls-\ls-\ls}
\vertex{8}{12}{12}{5.4}{-\ls-\ls-\ls}
\draw[-stealth, thick] (v2) -- node[anchor=east] {$\nEtJobS_{4,1}$} (v5);
\draw[-stealth, thick] (v3) -- node[anchor=east] {$\nEtJobS_{3,1}$} (v6);
\draw[-stealth, thick] (v3) -- node[anchor=west] {$\nEtJobS_{1,1}$} (v7);
\draw[-stealth, thick] (v4) -- node[anchor=west] {$\nEtJobS_{1,1}$} (v8);
\vertex{9}{12}{12}{-4}{-\ls-\ls-\ls-\ls}
\vertex{10}{14}{14}{0}{-\ls-\ls-\ls-\ls}
\vertex{11}{16}{16}{4}{-\ls-\ls-\ls-\ls}
\draw[-stealth, thick] (v5) -- node[anchor=east] {$\nEtJobS_{1,1}$} (v9);
\draw[-stealth, thick] (v6) -- node[anchor=west] {$\nEtJobS_{1,1}$} (v9);
\draw[-stealth, thick] (v7) -- node[anchor=west] {$\nEtJobS_{3,1}$} (v10);
\draw[-stealth, thick] (v8) -- node[anchor=west] {$\nEtJobS_{4,1}$} (v11);

\end{tikzpicture}
    \caption{Schedule graph for the example instance given in Table~\ref{table:sans_error_table} generated by SGA-ME using P-FP-EDF (non-work-conserving policy). Notice the two distinct arcs labeled $\nEtJobS_{3,1}$ leaving vertex $\nVert[1]$. The instance is proven to be schedulable.}
    \label{fig:sans_error_graph_mine}
\end{figure}

\begin{figure}[!hbt]
    % \includesvg[width=0.48\textwidth]{img/SANS_error/graph_wrong.svg}
    \centering
    \scriptsize
\newcommand{\vertex}[5] { % id, eft, lft, csx, csy
 	\node[ellipse, draw, thick, minimum size=0.1cm] (v#1) at (#4,#5) {$v_{#1}: [#2,#3]$};
}\newcommand{\redvertex}[5] { % id, eft, lft, csx, csy
 	\node[ellipse, draw, thick, minimum size=0.1cm, draw=red] (v#1) at (#4,#5) {$v_{#1}: [#2,#3]$};
}
\newcommand{\ls}{2}
\begin{tikzpicture}[scale=0.8, draw=black, text=black]
\vertex{0}{0}{0}{0}{0}
\vertex{1}{1}{8}{0}{-\ls}
\draw[-stealth, thick] (v0) -- node[anchor=west] {$\nEtJobS_{2,1}$} (v1);
\vertex{2}{3}{4}{-4}{-\ls-\ls}
\vertex{3}{7}{10}{0}{-\ls-\ls}
\vertex{4}{12}{12}{4}{-\ls-\ls}
\draw[-stealth, thick] (v1) -- node[anchor=east] {$\nEtJobS_{3,1}$~~~} (v2);
\draw[-stealth, thick] (v1) -- node[anchor=west] {$\nEtJobS_{4,1}$} (v3);
\draw[-stealth, thick] (v1) -- node[anchor=west] {~~$\nEtJobS_{1,1}$} (v4);
\vertex{5}{7}{8}{-5.4}{-\ls-\ls-\ls}
\vertex{6}{9}{10}{-1.8}{-\ls-\ls-\ls}
\vertex{7}{12}{12}{1.8}{-\ls-\ls-\ls}
\vertex{8}{16}{16}{5.4}{-\ls-\ls-\ls}
\draw[-stealth, thick] (v2) -- node[anchor=east] {$\nEtJobS_{4,1}$} (v5);
\draw[-stealth, thick] (v3) -- node[anchor=east] {$\nEtJobS_{3,1}$} (v6);
\draw[-stealth, thick] (v3) -- node[anchor=west] {$\nEtJobS_{1,1}$} (v7);
\draw[-stealth, thick] (v4) -- node[anchor=west] {$\nEtJobS_{4,1}$} (v8);
\vertex{9}{12}{12}{-4}{-\ls-\ls-\ls-\ls}
\vertex{10}{14}{14}{0}{-\ls-\ls-\ls-\ls}
\redvertex{11}{18}{18}{4}{-\ls-\ls-\ls-\ls}
\draw[-stealth, thick] (v5) -- node[anchor=east] {$\nEtJobS_{1,1}$} (v9);
\draw[-stealth, thick] (v6) -- node[anchor=west] {$\nEtJobS_{1,1}$} (v9);
\draw[-stealth, thick] (v7) -- node[anchor=west] {$\nEtJobS_{3,1}$} (v10);
\draw[-stealth, thick, draw=red, text=red] (v8) -- node[anchor=west] {$\nEtJobS_{3,1}$} (v11);

\end{tikzpicture}
    \caption{Schedule graph for the example instance given in Table~\ref{table:sans_error_table} generated by SGA-SE \cite{SANS} using P-FP-EDF (non-work-conserving policy). The generation yields a deadline miss depicted by red color.}
    \label{fig:sans_error_graph_their}
\end{figure}
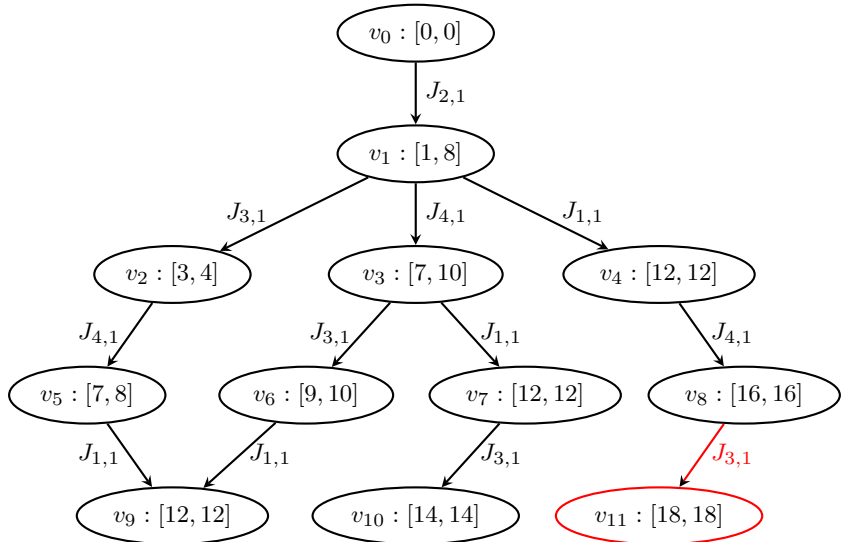

The complete schedule graph generated by SGA-ME using the P-FP-EDF policy can be seen in \figureautorefname{}~\ref{fig:sans_error_graph_mine}.
It is worth noticing that a vertex can have two outgoing arcs with the same job label, as in the case of vertex $\nVert[1]$. Note that this happens only for non-work-conserving policies.
On the contrary, the schedule graph generated by SGA-SE is depicted in \figureautorefname{}~\ref{fig:sans_error_graph_their}. It can be seen that it yields a deadline miss while the schedule graph generated by SGA-ME does not. 

We shall now explain in detail how the algorithms SGA-ME and SGA-SE differ in the process of the schedule graph generation on the problem instance described in Table~\ref{table:sans_error_table} using the P-FP-EDF policy. 

Throughout the entire schedule graph generation process, the only critical job may be job $\nEtJob{1}{1}$ as it is the only job with priority $p = 0$. Therefore, $\nVertCritTime = \nEtJobD{1}{1} - \nEtJobCMax{1}{1} = 12 - 2 = 10$ and $\nVertCritJob = \nEtJob{1}{1}$ for all nodes $\nVert$, where $\nEtJob{1}{1}$ is unfinished.

Both algorithms begin by scheduling job $\nEtJob{2}{1}$ as it is the only certainly released job at time $t = 0$ and it also does not violate the critical time because $\nEtJobCMax{2}{1} + t \leq \nVertCritTime[0]$, i.e., $8 + 0 \leq 10$. In both cases, the expansion phase of $\nVertLvl[0]$ results in a single new vertex $\nVert[1]$ with $\nVertEft[1] = 1$ and $\nVertLft[1] = 8$.

Where the two algorithms differ is during the expansion of vertex $\nVert[1]$. Here, job $\nEtJob{3}{1}$ is eligible at times $[1,2]$ and job $\nEtJob{4}{1}$ is eligible at times $[3,6]$. Note that $\nEtJob{3}{1}$ is not eigible at times $[3,6]$ because $\nEtJob{4}{1}$ has higher priority. Both approaches expand vertex $\nVert[1]$ by creating new vertices $\nVert[2]$ and $\nVert[3]$. However, the difference lies in $\nVert[4]$. SGA-ME finds that job $\nEtJob{3}{1}$ is also eligible at times $[7,8]$. Indeed, if job $\nEtJob{2}{1}$ starts its execution at time 0 and finishes at time 7 or 8, then according to the P-FP-EDF policy, the only released job that does not cause a deadline miss for the highest priority job $\nEtJob{1}{1}$ is job $\nEtJob{3}{1}$ and thus should be scheduled next. Hence, SGA-ME finds that job $\nEtJob{3}{1}$ is eligible in vertex $\nVert[1]$ in the time interval $[7,8]$. Hence, vertex $\nVert[4]$ is the result of expanding $\nVert[1]$ with job $\nEtJob{3}{1}$ in time range $[7,8]$, which results in $v_4.\nEft = 7 + \nEtJobCMin{3}{1} = 9$ and $v_4.\nLft = 8 + \nEtJobCMax{3}{1} = 10$.
On the contrary, SGA-SE \cite{SANS} concludes that there are no eligible jobs in the time interval $[7,9]$ and job $\nEtJob{1}{1}$ is scheduled next at its release time $t = \nEtJobRMin{1}{1} = \nEtJobRMax{1}{1} = 10$. As mentioned, this is due to the fact that according to \cite{SANS}, once a job is eligible during the expansion of a single vertex, it cannot become eligible again. In other words, according to \cite{SANS}, no vertex can have two outgoing arcs with the same job label. Therefore, job $\nEtJob{1}{1}$ is scheduled next, which results in vertex $\nVert[4]$ with $v_4.\nEft = \nEtJobRMin{1}{1} + \nEtJobCMin{1}{1} = 12$ and $v_4.\nLft = \nEtJobRMax{1}{1} + \nEtJobCMax{1}{1} = 12$. 
This eventually results in vertices $\nVert[8]$ and $\nVert[11]$ having different time intervals $[\nVert[8].\nEft, \nVert[8].\nLft]$ and $[\nVert[11].\nEft, \nVert[11].\nLft]$ which causes a deadline miss only for SGA-SE \cite{SANS}.

To sum up, SGA-SE checks the eligibility of each job $\nEtJob{i}{j}$ only at time $t_E = max(\nVertEft, \nEtJobRMin{i}{j})$ as follows from Definition~5 in \cite{SANS}. This means that once a job stops being eligible, it cannot become eligible again and hence no vertex can have two outgoing arcs with the same job label. That is why we distinguish single eligibility (SE) and multiple eligibility (ME) in the names of the algorithms.

\section{Experimental Evaluation} \label{4_ET_benchmark}

We evaluate the described algorithms empirically on randomly generated instances. We compare the performance of SGA-ME\footnote{\url{https://github.com/redakez/ettt-scheduling}} written in Java and SGA-SE\footnote{\url{https://github.com/gnelissen/np-schedulability-analysis}} \cite{SANS} written in C++. The experiments were run on a system with a 3550Mhz CPU, 8 cores, 16 threads; with 16 GB DDR4. Each Java virtual machine ran on a single core and was restricted to 4 GB of memory. Recall that the source codes of our implementation and the generated instances are available online.

\subsection{Performance Comparison of SGA-ME and SGA-SE}\label{sec:experimenty}

Recall that SGA-ME and SGA-SE yield different results for non-work-conserving policies as shown in Section~\ref{app_A_sans_error}. Hence, we evaluate the performance of the schedulability analyses only for work-conserving policy, namely, EDF.

We generated datasets with different release jitter ratio $\nGenPercJitt$ and execution time variation ratio $\nGenPercCVar$, which are defined as follows: For a task $\nEtTask[i]$, $\nGenPercCVar = (\nEtTaskCMax[i] - \nEtTaskCMin[i]) / (\nEtTaskCMax[i] - 1)$ if $\nEtTaskCMax[i] > 1$, otherwise $\nGenPercCVar = 0$. Similarly, $\nGenPercJitt = (\nEtTaskRMax[i] - \nEtTaskRMin[i]) / \nEtTaskRMax[i]$ if $\nEtTaskRMax[i] > 0$, otherwise $\nGenPercJitt = 0$. In particular, if $\nGenPercJitt = \nGenPercCVar = 0$, then $\nEtTaskRMin[i] = \nEtTaskRMax[i]$ and $\nEtTaskCMin[i] = \nEtTaskCMax[i]$, for all $\nEtTask[i] \in \nEtTask$.

Hence, we generated the following three datasets: dataset $\nDataset{1}{s}$ with $\nGenPercJitt = \nGenPercCVar = 0$, dataset $\nDataset{2}{s}$ with $\nGenPercJitt = \nGenPercCVar = 0.3$ and dataset $\nDataset{3}{s}$ with $\nGenPercJitt = \nGenPercCVar = 0.6$. Each dataset contains 960 instances. All generated instances have utilization $U = 0.3$, which is calculated as $U = \sum_{i=1}^{n} \frac{\nEtTaskCMax[i]}{\nEtTaskPe[i]}$.

We generated the instances such that $\nRMaxEtTask[i] + \nCMaxEtTask[i] \leq \nDEtTask[i] \leq \nPeEtTask[i]$, for each task $\nEtTask[i]$, and hence it suffices to use hyperperiod  $\nHyperPe = lcm(\nPeEtTask[1],\dots,\nPeEtTask[n])$, where $lcm$ is the least common multiple, as the observation interval.

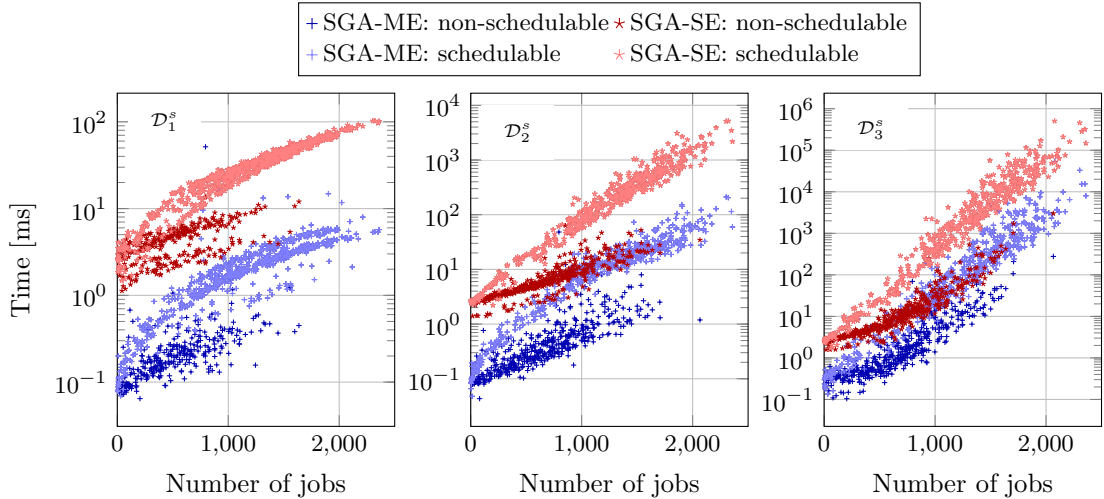
\begin{figure}[!hbt]
    \centering
    % \begin{minipage}[t]{0.24\textwidth}
    % \includegraphics[width=\linewidth]{tikz/sans/sans_D1_all.pdf}
    % \end{minipage}
    % \hfill
    % \begin{minipage}[t]{0.24\textwidth}
    % \includegraphics[width=\linewidth]{tikz/sans/sans_D1_schedulable.pdf}
    % \end{minipage}
    % \begin{minipage}[t]{0.24\textwidth}
    % \includegraphics[width=\linewidth]{tikz/sans/sans_D2_all.pdf}
    % \end{minipage}
    % \hfill
    % \begin{minipage}[t]{0.24\textwidth}
    % \includegraphics[width=\linewidth]{tikz/sans/sans_D2_schedulable.pdf}
    % \end{minipage}
    % \begin{minipage}[t]{0.24\textwidth}
    % \includegraphics[width=\linewidth]{tikz/sans/sans_D3_all.pdf}
    % \end{minipage}
    % \hfill
    % \begin{minipage}[t]{0.24\textwidth}
    % \includegraphics[width=\linewidth]{tikz/sans/sans_D3_schedulable.pdf}
    % \end{minipage}
    %\includegraphics[width=0.9\textwidth]{tikz/sans/sans_all.pdf}
\newcommand{\ClrFeasOur}{blue!50} % blue
\newcommand{\ClrInfeasOur}{blue!70!black} % blue!50
\newcommand{\ClrFeasSans}{red!50} % red
\newcommand{\ClrInfeasSans}{red!70!black} % red!50

\begin{tikzpicture}

	    \begin{groupplot}[group style={group size= 3 by 1, vertical sep=0.3cm},
				height=6cm,
				width=0.4\textwidth,
				xmin=0, xmax=2500,
				ymode=log,
			    ticklabel style={font=\scriptsize},
				ymajorgrids,
				xmajorgrids,
                legend image post style={scale=2.1},
				legend style={font=\scriptsize, legend columns=2},
				legend cell align={left},
		        transpose legend,
		        xlabel=Number of jobs
]

% =======================================================
	 \nextgroupplot[ytick={0.1, 1, 10, 100}, ylabel={Time [ms]},
]

				\addplot[only marks, \ClrInfeasOur, mark = +, mark size=1pt] table [x=Job_count, y=Time_ettt, col sep=space] {measurements/ettt_sans/ettt_sans_no_jitter.dat};
					\addplot[only marks, \ClrFeasOur, mark = +, mark size=1pt] table [x=Job_count, y=Time_ettt, col sep=space] {measurements/ettt_sans/ettt_sans_no_jitter_schedulable.dat};

					\addplot[only marks, \ClrInfeasSans, mark = star, mark size=1pt] table [x=Job_count, y=Time_nptest, col sep=space] {measurements/ettt_sans/ettt_sans_no_jitter.dat};
					\addplot[only marks, \ClrFeasSans, mark = star, mark size=1pt] table [x=Job_count, y=Time_nptest, col sep=space] {measurements/ettt_sans/ettt_sans_no_jitter_schedulable.dat};

\path[fill=white, fill opacity=0.25, text opacity=1] (axis cs: 100, 50) rectangle node {\footnotesize $\nDataset{1}{s}$} (axis cs: 750, 200);

% =======================================================
			 \nextgroupplot[ytick={0.1, 1, 10, 100, 1000, 10000}, legend pos=north west, legend to name=lgnd]

				\addplot[only marks, \ClrInfeasOur, mark = +, mark size=1pt] table [x=Job_count, y=Time_ettt, col sep=space] {measurements/ettt_sans/ettt_sans_low_jitter.dat};
				\addlegendentry{SGA-ME: non-schedulable}
					\addplot[only marks, \ClrFeasOur, mark = +, mark size=1pt] table [x=Job_count, y=Time_ettt, col sep=space] {measurements/ettt_sans/ettt_sans_low_jitter_schedulable.dat};
					\addlegendentry{SGA-ME: schedulable}

					\addplot[only marks, \ClrInfeasSans, mark = star, mark size=1pt] table [x=Job_count, y=Time_nptest, col sep=space] {measurements/ettt_sans/ettt_sans_low_jitter.dat};
					\addlegendentry{SGA-SE: non-schedulable}
					\addplot[only marks, \ClrFeasSans, mark = star, mark size=1pt] table [x=Job_count, y=Time_nptest, col sep=space] {measurements/ettt_sans/ettt_sans_low_jitter_schedulable.dat};
					\addlegendentry{SGA-SE: schedulable}

\path[fill=white, fill opacity=0.25, text opacity=1] (axis cs: 100, 1000) rectangle node {\footnotesize $\nDataset{2}{s}$} (axis cs: 750, 10000);

   \coordinate (c1) at (rel axis cs:0,1);
	\coordinate (c2) at (rel axis cs:1,1);% I moved this to the upper right corner

% =======================================================
			 \nextgroupplot[ytick={0.1, 1, 10, 100, 1000, 10000, 100000, 1000000}] 

					\addplot[only marks, \ClrInfeasOur, mark = +, mark size=1pt] table [x=Job_count, y=Time_ettt, col sep=space] {measurements/ettt_sans/ettt_sans_high_jitter.dat};
					
					\addplot[only marks, \ClrFeasOur, mark = +, mark size=1pt] table [x=Job_count, y=Time_ettt, col sep=space] {measurements/ettt_sans/ettt_sans_high_jitter_schedulable.dat};

					\addplot[only marks, \ClrInfeasSans, mark = star, mark size=1pt] table [x=Job_count, y=Time_nptest, col sep=space] {measurements/ettt_sans/ettt_sans_high_jitter.dat};
					
					\addplot[only marks, \ClrFeasSans, mark = star, mark size=1pt] table [x=Job_count, y=Time_nptest, col sep=space] {measurements/ettt_sans/ettt_sans_high_jitter_schedulable.dat};

\path[fill=white, fill opacity=0.25, text opacity=1] (axis cs: 100, 100000) rectangle node {\footnotesize $\nDataset{3}{s}$} (axis cs: 750, 1000000);

    \end{groupplot}

   \coordinate (c3) at ($(c1)!.5!(c2)$);
    \node[above] at (c3 |- current bounding box.north)
      {\pgfplotslegendfromname{lgnd}};

\end{tikzpicture}

    \caption{Solving times of SGA-ME and SGA-SE on datasets $\nDataset{1}{s}$, $\nDataset{2}{s}$ and $\nDataset{3}{s}$, respectively from left to right. Schedulable instances are depicted with light color, non-schedulable with dark color.}
    \label{fig:ettt_vs_sans_high_jitter}
\end{figure}

The solving times can be seen in \figureautorefname{}~\ref{fig:ettt_vs_sans_high_jitter}. Generally speaking, the larger the release jitter and execution time variation, the longer the solving time. Notice the increasing (logarithmic) scale on the vertical axis with each subsequent dataset. 
As can be seen in the charts, the schedulability analyses seem to have the same asymptotic behavior. The non-schedulable instances finish quicker because the algorithms do not need to compute the entire schedule graph but they finish as soon as they discover the first deadline miss.

\begin{figure}[!hbt]
    \centering
    \resizebox{0.5\textwidth}{!}{%
    \begin{tikzpicture}
    \begin{axis} [
      xtick = {1,2,3},
      xticklabels={$\nDataset{1}{s}$, $\nDataset{2}{s}$, $\nDataset{3}{s}$},
      ylabel={Speed-up},
      boxplot/draw direction=y,
      boxplot/variable width,
      ymin=0,ymax=40
    ]
    \addplot[boxplot] table[y index=0] {measurements/ettt_sans/ettt_sans_speedups.dat};
    \addplot[boxplot] table[y index=1] {measurements/ettt_sans/ettt_sans_speedups.dat};
    \addplot[boxplot] table[y index=2] {measurements/ettt_sans/ettt_sans_speedups.dat};
    \end{axis}
    \end{tikzpicture}%
    }
    
    \caption{Box plots of speed-ups for the datasets $\nDataset{1}{s}$, $\nDataset{2}{s}$ and $\nDataset{3}{s}$ between SGA-ME and SGA-SE.}
    \label{fig:ettt_vs_sans_speedups}
\end{figure}
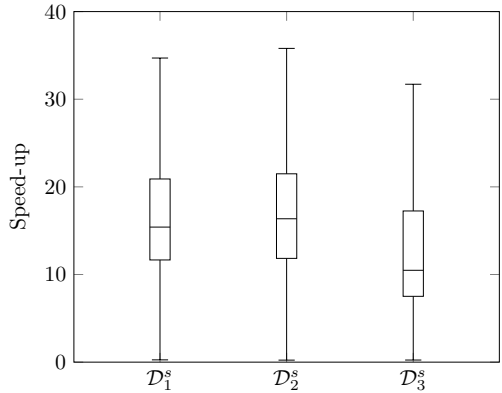

Additionally, it can be immediately seen that SGA-ME is substantially faster than SGA-SE, which is most likely caused by the new definitions of eligibility rules. \figureautorefname{}~\ref{fig:ettt_vs_sans_speedups} shows box plots of speed-ups of SGA-ME over SGA-SE, which we define as the solving time of SGA-SE divided by the solving time of SGA-ME. The average speed-ups for datasets $\nDataset{1}{s}$, $\nDataset{2}{s}$ and $\nDataset{3}{s}$ are 17.64, 17.16 and 12.31, respectively.

\section{Conclusion} \label{7_Conclusion}

This paper addresses the problem of schedulability analysis for non-preemptive tasks on a uniprocessor. 
We focused on existing schedulability analysis and improved the approach presented in \cite{SANS} such that it builds a schedule graph using new definitions of eligibility rules and thus is exact and sustainable for both work-conserving and enhanced formalization of non-work-conserving policies. Besides, we compared our algorithm to that of \cite{SANS} in terms of solving time, showing that our algorithm was faster by an order of magnitude.

The proposed algorithms based on schedule graph generation can be further accelerated by partial-order reduction as presented in \cite{ranjha2022partial}, which keeps the exactness of the schedulability analysis but introduces over-estimation in the worst-case response times.

The notion of the schedule graph presents an exceptionally valuable framework for integrating online scheduling of event-triggered jobs and offline schedule synthesis for time-triggered jobs. We already extended the algorithm presented in this paper by initial study on the integration of the offline schedule synthesis on a uniprocessor \cite{jaros2022combination} as well as on dedicated resources and precedences \cite{halaska2023combination}.

\backmatter

\bmhead{Acknowledgements}

This work was co-funded by the European Union under the project ROBOPROX (reg. no. CZ.02.01.01/00/22\_008/0004590) and by the Grant Agency of the Czech Republic under the Project GACR 22-31670S.

\begin{appendices}

\section{Scheduling Policies}\label{sec:policies}

This section describes scheduling policies used by an online scheduler. 

\subsection{Work-Conserving Scheduling Policies}

We start with the work-conserving scheduling policies, which are those that never idle when there is at least one released unfinished job.

\subsubsection{Earliest Deadline First (EDF)}

We have already introduced the EDF scheduling policy in Section \ref{2_formal_desc} as an example. A pseudocode of the EDF scheduling policy can be seen in Algorithm \ref{EDF_policy}. Note that this policy does not make use of the priority values $\nPr$.

\begin{algorithm}
\footnotesize
    \caption{EDF Scheduling Policy}\label{EDF_policy}
    % \hspace*{\algorithmicindent} \textbf{Input:} Time $t$ and applicable jobs $\nAppJobs$ \\
    % \hspace*{\algorithmicindent} \textbf{Output:} Selected job $\nEtJobS_e$ or $null$\\
    \begin{algorithmic}[1]
    \Function{EDF\_POLICY}{$t$,$\nAppJobs$}
    \State \textbf{$\nEtJobS^R$} $\gets$ subset of $\nAppJobs$ with released jobs only
    \If{$\nEtJobS^R = \emptyset$}
    \State \textbf{return} null
    \EndIf
    \State \textbf{return} job from $\nEtJobS^R$ with the lowest deadline
    \EndFunction
    \end{algorithmic}
\end{algorithm}

\subsubsection{Fixed Priority - EDF (FP-EDF)}

The FP-EDF is a version of the EDF policy that uses priority value $\nPr$ of the jobs. A pseudocode of the FP-EDF policy can be seen in Algorithm~\ref{FP-EDF_policy}. The line \ref{FP-EDF_policy_sort_line} returns the job with the lowest value of $\nPr$ (the highest priority), breaking the ties by taking the job with the lowest deadline. %If multiple released jobs $\nEtJobS$ have the same lowest value $\nPr$, the one with the lowest deadline is chosen.

\begin{algorithm}
\footnotesize
    \caption{FP-EDF Scheduling Policy}\label{FP-EDF_policy}
    % \hspace*{\algorithmicindent} \textbf{Input:} Time $t$ and applicable jobs $\nAppJobs$ \\
    % \hspace*{\algorithmicindent} \textbf{Output:} Selected job $\nEtJobS_e$ or $null$\\
    \begin{algorithmic}[1]
    \Function{FP-EDF\_POLICY}{$t$,$\nAppJobs$}
    \State \textbf{$\nEtJobS^R$} $\gets$ subset of $\nAppJobs$ with released jobs only
    \If{$\nEtJobS^R = \emptyset$}
    \State \textbf{return} null
    \EndIf
    \State \textbf{return} job from $\nEtJobS^R$ with the lowest $\nPr$ value, then the lowest deadline \label{FP-EDF_policy_sort_line}
    \EndFunction
    \end{algorithmic}
\end{algorithm}

By using priority values $\nPr$, the resulting schedule may execute jobs with higher priority earlier. Another useful property of the FP-EDF policy is that it can be easily changed to the EDF policy by setting the priority of all jobs to the same value. It can also be changed to \textit{fixed priority policy} (FP policy), which schedules jobs based only on priority values $\nPr$. This change is done by setting unique priority values $\nPr$ to each task. %Due to these advantages, we use this as the go-to scheduling policy in the following policies.

\subsection{Non-Work-Conserving Policies}

While work-conserving schedulers always pick one of the released jobs, non-work-conserving schedulers may let the resource stay idle until some other job is released. Non-work-conserving policies may thus avoid some deadline misses and lead to higher schedulability. We will now describe three non-work-conserving policies that we also evaluate experimentally.

\subsubsection{Precautious FP-EDF (P-FP-EDF)}

Under the P-FP-EDF scheduling policy, a job with a priority different than the highest one cannot be scheduled if it may cause a deadline miss for some \textit{critical job}. The critical job is the applicable job that has the lowest $\nRMax$ out of all jobs for which $p=0$ \cite{PRM}. Recall that $\nPr = 0$ is the highest priority.

In Algorithm~\ref{PRM_policy}, the critical job is denoted as $\nCritJob$. The latest time by which the picked job must be finished is called \textit{critical time} and is denoted as $\nCritTime$. Released, applicable jobs that do not cause a deadline miss to $\nCritJob$ are called viable jobs and are denoted as $\nEtJobS^V$.

Unlike in \cite{PRM}, our version of the P-FP-EDF policy follows the FP-EDF policy for jobs that can be scheduled without causing a deadline miss for the critical job. 

\begin{algorithm}
\footnotesize
    \caption{P-FP-EDF Scheduling Policy} \label{PRM_policy}
    % \hspace*{\algorithmicindent} \textbf{Input:} Time $t$ and applicable jobs $\nAppJobs$ \\
    % \hspace*{\algorithmicindent} \textbf{Output:} Selected job $\nEtJobS_e$ or $null$\\
    \begin{algorithmic}[1]
    \Function{P-FP-EDF\_POLICY}{$t$,$\nAppJobs$}
    \State \textbf{$\nCritJob$} $\gets$ job from $\nAppJobs$ with $\nPr = 0$ and the lowest $\nRMax$
    \If{$\nCritJob$ is $null$}
    \State \textbf{return} FP-EDF\_POLICY($t$,$\nAppJobs$)
    \EndIf
    \State \textbf{$\nCritTime$} $\gets \nCritJob.\nD - \nCritJob.\nCMax$
    \State \textbf{$\nEtJobS^R$} $\gets$ subset of $\nAppJobs$ with released jobs only
    \State \textbf{$\nEtJobS^V$} $\gets \{\nEtJob{i}{j} | \nEtJob{i}{j} \in \nEtJobS^R \wedge ( t+\nEtJobCMax{i}{j} \leq \nCritTime\ \vee \nEtJob{i}{j} = \nCritJob )\}$
    \If{$\nEtJobS^V = \emptyset$}
    \State \textbf{return} $null$
    \EndIf
    \State \textbf{return} job from $\nEtJobS^V$ with the lowest $\nPr$ value, then the lowest deadline
    \EndFunction
    
    \end{algorithmic}
\end{algorithm}

\subsubsection{Critical Point (CP)}

CP is a scheduling policy similar to P-FP-EDF with the main difference being that the critical job is the one with the lowest deadline. 
%follows the FP-EDF scheduling policy but does not schedule a job if it will cause a deadline miss to an unfinished job with the lowest deadline. This job is denoted as $\nCritJob$ and is again called a critical job. 
The pseudocode of the CP scheduling policy can be seen in Algorithm \ref{CP_policy}.
Notice that the P-FP-EDF policy may not have a critical job while the CP policy always has a critical job if $\nAppJobs$ is not empty.

\begin{algorithm}
\footnotesize
    \caption{CP Scheduling Policy} \label{CP_policy}
    % \hspace*{\algorithmicindent} \textbf{Input:} Time $t$ and applicable jobs $\nAppJobs$ \\
    % \hspace*{\algorithmicindent} \textbf{Output:} Selected job $\nEtJobS_e$ or $null$\\
    \begin{algorithmic}[1]
    \Function{CP\_POLICY}{$t$,$\nAppJobs$}
    \If{$\nAppJobs = \emptyset$}
    \State \textbf{return} null
    \EndIf
    \State \textbf{$\nCritJob$} $\gets$ job from $\nAppJobs$ with the lowest deadline \label{CP_policy_diff}
    \State \textbf{$\nCritTime$} $\gets \nCritJob.\nD - \nCritJob.\nCMax$ 
    \State \textbf{$\nEtJobS^R$} $\gets$ subset of $\nAppJobs$ with released jobs only
    \State \textbf{$\nEtJobS^V$} $\gets \{\nEtJob{i}{j} | \nEtJob{i}{j} \in \nEtJobS^R \wedge ( t+\nEtJobCMax{i}{j} \leq \nCritTime\ \vee \nEtJob{i}{j} = \nCritJob )\}$
    \If{$\nEtJobS^V = \emptyset$}
    \State \textbf{return} $null$
    \EndIf
    \State \textbf{return} job from $\nEtJobS^V$ with the lowest $\nPr$ value, then the lowest deadline
    \EndFunction
    \end{algorithmic}
\end{algorithm}

%This policy is similar to P-FP-EDF in the sense that both define a critical job and no other job can be executed if it will cause a deadline miss for the critical job. The only difference between the two policies is how the critical job is selected and that the P-FP-EDF policy may not have a critical job while the CP policy always has a critical job if $\nAppJobs$ is not empty.

\subsubsection{Critical Window (CW)}

%Right now, the policy is implemented differently than in \cite{SANS} and may be re-implemented later to either suit the description in \cite{SANS} or changed in some other way.

CW differs from CP by determining its critical time using all applicable jobs instead of a single job. Its pseudocode can be seen in Algorithm \ref{MCW_policy}. Note that on line \ref{lbl:orderedset}, $\nEtJobS^S$ is an ordered set of jobs, which is relevant in the cycle on line \ref{lbl:forcycle}. This scheduling policy is a version of CW-EDF from \cite{SANS} modified to fit into our solution approach.

\begin{algorithm}
\footnotesize
    \caption{CW Scheduling Policy} \label{MCW_policy}
    % \hspace*{\algorithmicindent} \textbf{Input:} Time $t$ and applicable jobs $\nAppJobs$ \\
    % \hspace*{\algorithmicindent} \textbf{Output:} Selected job $\nEtJobS_e$ or $null$\\
    \begin{algorithmic}[1]
    \Function{CW\_POLICY}{$t$,$\nAppJobs$}
    \If{$\nAppJobs = \emptyset$}
    \State \textbf{return} null
    \EndIf
    \State \textbf{$\nEtJobS^S$} $\gets$ $\nAppJobs$ sorted by $\nD$ in descending order\label{lbl:orderedset}
    \State \textbf{$\nCritTime$} $\gets \infty$
    \For {\textbf{each} $\nEtJob{i}{j} \in \nEtJobS^S$}\label{lbl:forcycle}
    %\If{$\nEtJobD{i}{j} < \nCritTime$}
    \State \textbf{$\nCritTime$} $\gets \min\{\nCritTime, \nEtJobD{i}{j}\} - \nEtJobCMax{i}{j}$
    %\EndIf
    %\If{$\nEtJobD{i}{j} \geq \nCritTime$}
    %\State \textbf{$\nCritTime$} $\gets \nCritTime - \nEtJobCMax{i}{j}$ 
    %\EndIf
    \EndFor
    \State \textbf{$\nCritJob$} $\gets$ the last job in $\nEtJobS^S$
    \State \textbf{$\nEtJobS^R$} $\gets$ subset of $\nAppJobs$ with released jobs only
    \State \textbf{$\nEtJobS^V$} $\gets \{\nEtJob{i}{j} | \nEtJob{i}{j} \in \nEtJobS^R \wedge ( t+\nEtJobCMax{i}{j} \leq \nCritTime\ \vee \nEtJob{i}{j} = \nCritJob )\}$
    \If{$\nEtJobS^V = \emptyset$}
    \State \textbf{return} $null$
    \EndIf
    \State \textbf{return} job from $\nEtJobS^V$ with the lowest $\nPr$ value, then the lowest deadline
    \EndFunction
    \end{algorithmic}
\end{algorithm}

\subsection{Schedulability of Different Policies}

Next, we evaluate the schedulability of the three described non-work-conserving policies. The parameters affecting the generation of random instances are explained in Section~\ref{sec:experimenty}. We generated 3 datasets $\nDataset{1}{p}$, $\nDataset{2}{p}$ and $\nDataset{3}{p}$ with $\nGenPercJitt$ and $\nGenPercCVar$ set to 0, 0.3 and 0.6, respectively. Each dataset contains 200 instances for each utilization $U \in \{0.1, 0.2, \dots, 0.9\}$, i.e., 1800 instances for each dataset. 
All task priorities are set to zero.

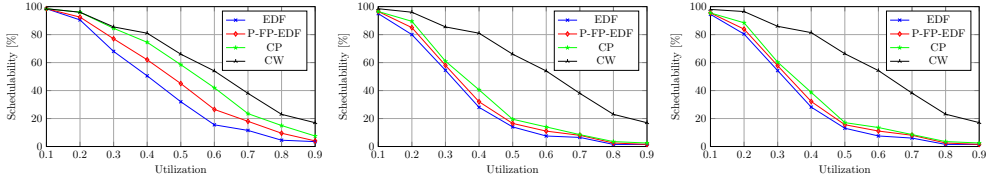
\begin{figure*}
    \centering
    
    \resizebox{0.325\textwidth}{!}{%
    \begin{tikzpicture}
    \begin{axis}[
      width=100mm,height=60mm,
      xmin=0.1,ymin=0,xmax=0.9,ymax=1,
      xlabel={Utilization},ylabel={Schedulability [\%]},
      yticklabel={\pgfmathparse{\tick*100}\pgfmathprintnumber{\pgfmathresult}},
      legend pos=north east,
      grid=both,
      grid style={line width=.1pt, draw=gray!50, opacity = 0},
      major grid style={line width=.2pt,draw=gray!70, opacity = 1},
    ]
    \addplot[blue, mark=x]
    table[x=utilization,y=no_iip] {measurements/ettt/ettt_no_jitter_schedulability.dat};
    \addlegendentry{EDF}
    \addplot[red, mark=diamond]
    table[x=utilization,y=PRM] {measurements/ettt/ettt_no_jitter_schedulability.dat};
    \addlegendentry{P-FP-EDF}
    \addplot[green, mark=star]
    table[x=utilization,y=CP] {measurements/ettt/ettt_no_jitter_schedulability.dat};
    \addlegendentry{CP}
    \addplot[black, mark=Mercedes star]
    table[x=utilization,y=CW] {measurements/ettt/ettt_no_jitter_schedulability.dat};
    \addlegendentry{CW}
    \end{axis}
    \end{tikzpicture}%
    }
    \resizebox{0.325\textwidth}{!}{%
    \begin{tikzpicture}
    \begin{axis}[
      width=100mm,height=60mm,
      xmin=0.1,ymin=0,xmax=0.9,ymax=1,
      xlabel={Utilization},ylabel={Schedulability [\%]},
      yticklabel={\pgfmathparse{\tick*100}\pgfmathprintnumber{\pgfmathresult}},
      legend pos=north east,
      grid=both,
      grid style={line width=.1pt, draw=gray!50, opacity = 0},
      major grid style={line width=.2pt,draw=gray!70, opacity = 1},
    ]
    \addplot[blue, mark=x]
    table[x=utilization,y=no_iip] {measurements/ettt/ettt_low_jitter_schedulability.dat};
    \addlegendentry{EDF}
    \addplot[red, mark=diamond]
    table[x=utilization,y=PRM] {measurements/ettt/ettt_low_jitter_schedulability.dat};
    \addlegendentry{P-FP-EDF}
    \addplot[green, mark=star]
    table[x=utilization,y=CP] {measurements/ettt/ettt_low_jitter_schedulability.dat};
    \addlegendentry{CP}
    \addplot[black, mark=Mercedes star]
    table[x=utilization,y=CW] {measurements/ettt/ettt_low_jitter_schedulability.dat};
    \addlegendentry{CW}
    \end{axis}
    \end{tikzpicture}%
    }
    \resizebox{0.325\textwidth}{!}{%
    \begin{tikzpicture}
    \begin{axis}[
      width=100mm,height=60mm,
      xmin=0.1,ymin=0,xmax=0.9,ymax=1,
      xlabel={Utilization},ylabel={Schedulability [\%]},
      yticklabel={\pgfmathparse{\tick*100}\pgfmathprintnumber{\pgfmathresult}},
      legend pos=north east,
      grid=both,
      grid style={line width=.1pt, draw=gray!50, opacity = 0},
      major grid style={line width=.2pt,draw=gray!70, opacity = 1},
    ]
    \addplot[blue, mark=x]
    table[x=utilization,y=no_iip] {measurements/ettt/ettt_high_jitter_schedulability.dat};
    \addlegendentry{EDF}
    \addplot[red, mark=diamond]
    table[x=utilization,y=PRM] {measurements/ettt/ettt_high_jitter_schedulability.dat};
    \addlegendentry{P-FP-EDF}
    \addplot[green, mark=star]
    table[x=utilization,y=CP] {measurements/ettt/ettt_high_jitter_schedulability.dat};
    \addlegendentry{CP}
    \addplot[black, mark=Mercedes star]
    table[x=utilization,y=CW] {measurements/ettt/ettt_high_jitter_schedulability.dat};
    \addlegendentry{CW}
    \end{axis}
    \end{tikzpicture}%
    }
    
    \caption{Schedulability of policies on datasets $\nDataset{1}{p}$, $\nDataset{2}{p}$ and $\nDataset{3}{p}$, respectively from left to right.}
    \label{fig:ettt_schedulability}
\end{figure*}

Schedulability of policies for all datasets can be seen in \figureautorefname{}~\ref{fig:ettt_schedulability}. The schedulability of all policies except the CW policy decreases with increasing release jitter and execution time variation.

Besides the schedulability comparison in graphs, we note that the EDF policy runs the fastest as it uses simplified rules of schedule graph generation. Solving times of the CP and P-FP-EDF policies are about the same. This is likely because both policies search through the entire set of applicable jobs to find the critical job and the critical time. The CW policy is the slowest as it has to sort the set of applicable jobs and then process each job in order to find the critical job and the critical time.

\section{Implementation of the Next Nodes Routine}\label{sec:nextnodesimpl}
The Algorithm \ref{exp_phase_edffp_complex2} presents more detailed pseudocode for function NEXT\_NODES for work-conserving policies presented in Algorithm~\ref{exp_phase_edffp}.

%\begin{fullwidth}[width=\linewidth+1cm,leftmargin=-1cm,rightmargin=-1cm]
\begin{algorithm}
\footnotesize
    \caption{NEXT\_NODES for Work-Conserving Policies}\label{exp_phase_edffp_complex2}
    % \hspace*{\algorithmicindent} \textbf{Input:} vertex $\nVert$ \\
    % \hspace*{\algorithmicindent} \textbf{Output:} set of vertices $\nVertLvlN[new]$ \\
    \begin{algorithmic}[1]
    
    \Function{NEXT\_NODES\_CONSERVING}{$\nVert$}
    \If {$\nVertAppJobs = \emptyset$}
    \State \textbf{return} $\emptyset$
    \EndIf
    \State \textbf{$\nVertLvlN[new] \gets$} $\emptyset$
    \State \textbf{$\nLftExt \gets min\{t \mid t \geq \nVertLft \wedge \nCeJob[t] \; exists\}$}
    \State \textbf{$CE \gets$} null
    \State \textbf{$PE \gets$} $\emptyset$
    %\For {$t=\nVertEft$ ; $t \leq \nLftExt+1$ ; $t$\texttt{++}} %some other notation?
    
    \For {$t = \nVertEft ~\textbf{to}~ \nLftExt+1$}
    
    \If {$t = \nLftExt + 1$}
    \State \textbf{$\nCeJob[t] \gets$} newly created job st. $\nCeJob[t].\nD = \nCeJob[t].\nPr = -1$ \label{exp_phase_edffp_complex2_job}
    \Else
    \State \textbf{$\nCeJob[t] \gets$} certainly-eligible job at time $t$ \label{exp_phase_edffp_complex2_null}
    \EndIf
    
    \If {$\nCeJob[t] \neq null \wedge \nCeJob[t] \neq CE$}
    \If {$CE \neq null$}
    \State \textbf{$\nVert[k] \gets$} EXPAND($\nVert,CE,max(\nVertEft,CE.\nRMin),t-1$)
    \State $\nVertLvlN[new] \gets \nVertLvlN[new] \cup \{\nVert[k]\}$
    \EndIf
    \If {$\nCeJob[t] \in PE$}
    \State $PE \gets PE \setminus \{\nCeJob[t]\}$
    \EndIf
    \State $CE \gets \nCeJob[t]$
    \For {\textbf{each} $\nEtJobS \in PE$}
    \If {$\nEtJobS$ is not possibly-eligible at time $t$}
    \State \textbf{$\nVert[k] \gets$} EXPAND($\nVert,\nEtJobS,max(\nVertEft,\nEtJobS.r^{min}),t-1$)
    \State $\nVertLvlN[new] \gets \nVertLvlN[new] \cup \{\nVert[k]\}$
    \State $PE \gets PE \setminus \{\nEtJobS\}$
    \EndIf
    \EndFor
    \EndIf
    
    \For {\textbf{each} $\nEtJobS \in \nAppJobs$}
    \If {$\nEtJobS$ is possibly-eligible at time $t$ $\wedge$ $\nEtJobS \not\in PE$}
    \State $PE \gets PE \cup \{\nEtJobS\}$
    \EndIf
    \EndFor
    
    \EndFor
    
    \State \textbf{return} $\nVertLvlN[new]$
    \EndFunction
    \end{algorithmic}
\end{algorithm}
%\end{fullwidth}

This pseudocode goes through each integer time $t \in [\nVertEft, \nLftExt+1]$ and examines how the certainly and possibly released jobs change. Variable CE represents the current certainly-eligible job, and variable PE represents the current set of possibly-eligible jobs. For each time $t \in [\nVertEft, \nLftExt]$, the algorithm finds the certainly-eligible job $\nCeJob[t]$ and, if needed, changes job CE and jobs in PE according to $\nCeJob[t]$. Lastly, it adds new possibly-eligible jobs to PE. Note that on line \ref{exp_phase_edffp_complex2_null}, $\nCeJob[t]$ is null if the certainly released job does not exist. 

When changing job CE or removing some job from PE, the algorithm expands vertex $\nVert$. This is because when a job is no longer possibly or certainly-eligible, we know the time interval during which the job was eligible and hence we know the parameters needed to create new vertex $\nVert[k]$.

During the last iteration of the algorithm, i.e., $t = \nLftExt+1$, the certainly-eligible job is a dummy job that has higher $\nPolicy$-priority than any other applicable job in $\nVertAppJobs$. 
This dummy job is created on line \ref{exp_phase_edffp_complex2_job}. For the FP-EDF and EDF policies, this is realized by setting $\nCeJob[t].\nD = \nCeJob[t].\nPr = -1$.
Thanks to this dummy job, all the jobs left in variables CE and PE are removed, which means they expand vertex $\nVert$.

In our final implementation, we further improve the Algorithm~\ref{exp_phase_edffp_complex2} such that it does not iterate over all times $t \in [\nVertEft, \nLftExt+1]$, but only over times $t \in \{t \mid \nEtJob{i}{j} \in \nAppJobs \wedge (\nEtJobRMin{i}{j} = t \vee \nEtJobRMax{i}{j} = t)\} \cup \{\nLftExt+1\}$.

\end{appendices}

\bibliography{sn-article}


\begin{thebibliography}{22}
\ifx \bisbn   \undefined \def \bisbn  #1{ISBN #1}\fi
\ifx \binits  \undefined \def \binits#1{#1}\fi
\ifx \bauthor  \undefined \def \bauthor#1{#1}\fi
\ifx \batitle  \undefined \def \batitle#1{#1}\fi
\ifx \bjtitle  \undefined \def \bjtitle#1{#1}\fi
\ifx \bvolume  \undefined \def \bvolume#1{\textbf{#1}}\fi
\ifx \byear  \undefined \def \byear#1{#1}\fi
\ifx \bissue  \undefined \def \bissue#1{#1}\fi
\ifx \bfpage  \undefined \def \bfpage#1{#1}\fi
\ifx \blpage  \undefined \def \blpage #1{#1}\fi
\ifx \burl  \undefined \def \burl#1{\textsf{#1}}\fi
\ifx \doiurl  \undefined \def \doiurl#1{\url{https://doi.org/#1}}\fi
\ifx \betal  \undefined \def \betal{\textit{et al.}}\fi
\ifx \binstitute  \undefined \def \binstitute#1{#1}\fi
\ifx \binstitutionaled  \undefined \def \binstitutionaled#1{#1}\fi
\ifx \bctitle  \undefined \def \bctitle#1{#1}\fi
\ifx \beditor  \undefined \def \beditor#1{#1}\fi
\ifx \bpublisher  \undefined \def \bpublisher#1{#1}\fi
\ifx \bbtitle  \undefined \def \bbtitle#1{#1}\fi
\ifx \bedition  \undefined \def \bedition#1{#1}\fi
\ifx \bseriesno  \undefined \def \bseriesno#1{#1}\fi
\ifx \blocation  \undefined \def \blocation#1{#1}\fi
\ifx \bsertitle  \undefined \def \bsertitle#1{#1}\fi
\ifx \bsnm \undefined \def \bsnm#1{#1}\fi
\ifx \bsuffix \undefined \def \bsuffix#1{#1}\fi
\ifx \bparticle \undefined \def \bparticle#1{#1}\fi
\ifx \barticle \undefined \def \barticle#1{#1}\fi
\bibcommenthead
\ifx \bconfdate \undefined \def \bconfdate #1{#1}\fi
\ifx \botherref \undefined \def \botherref #1{#1}\fi
\ifx \url \undefined \def \url#1{\textsf{#1}}\fi
\ifx \bchapter \undefined \def \bchapter#1{#1}\fi
\ifx \bbook \undefined \def \bbook#1{#1}\fi
\ifx \bcomment \undefined \def \bcomment#1{#1}\fi
\ifx \oauthor \undefined \def \oauthor#1{#1}\fi
\ifx \citeauthoryear \undefined \def \citeauthoryear#1{#1}\fi
\ifx \endbibitem  \undefined \def \endbibitem {}\fi
\ifx \bconflocation  \undefined \def \bconflocation#1{#1}\fi
\ifx \arxivurl  \undefined \def \arxivurl#1{\textsf{#1}}\fi
\csname PreBibitemsHook\endcsname

%%% 1
\bibitem[\protect\citeauthoryear{Ekberg}{2020}]{ekberg2020rate}
\begin{bchapter}
\bauthor{\bsnm{Ekberg}, \binits{P.}}:
\bctitle{Rate-monotonic schedulability of implicit-deadline tasks is {NP}-hard
  beyond {L}iu and {L}ayland’s bound}.
In: \bbtitle{2020 IEEE Real-Time Systems Symposium (RTSS)},
pp. \bfpage{308}--\blpage{318}
(\byear{2020}).
\bcomment{IEEE}
\end{bchapter}
\endbibitem

%%% 2
\bibitem[\protect\citeauthoryear{Baruah and
  Goossens}{2004}]{baruah2004scheduling}
\begin{botherref}
\oauthor{\bsnm{Baruah}, \binits{S.}},
\oauthor{\bsnm{Goossens}, \binits{J.}}:
Scheduling real-time tasks: Algorithms and complexity.
Handbook of scheduling: Algorithms, models, and performance analysis
\textbf{3}
(2004)
\end{botherref}
\endbibitem

%%% 3
\bibitem[\protect\citeauthoryear{Park and Cho}{2004}]{park2004feasibility}
\begin{barticle}
\bauthor{\bsnm{Park}, \binits{M.}},
\bauthor{\bsnm{Cho}, \binits{Y.}}:
\batitle{Feasibility analysis of hard real-time periodic tasks}.
\bjtitle{Journal of Systems and Software}
\bvolume{73}(\bissue{1}),
\bfpage{89}--\blpage{100}
(\byear{2004})
\end{barticle}
\endbibitem

%%% 4
\bibitem[\protect\citeauthoryear{Baruah and Burns}{2006}]{SUS}
\begin{bchapter}
\bauthor{\bsnm{Baruah}, \binits{S.}},
\bauthor{\bsnm{Burns}, \binits{A.}}:
\bctitle{Sustainable scheduling analysis}.
In: \bbtitle{2006 27th IEEE International Real-Time Systems Symposium
  (RTSS'06)},
pp. \bfpage{159}--\blpage{168}
(\byear{2006}).
\bcomment{IEEE}
\end{bchapter}
\endbibitem

%%% 5
\bibitem[\protect\citeauthoryear{Davis et~al.}{2007}]{davis2007controller}
\begin{barticle}
\bauthor{\bsnm{Davis}, \binits{R.I.}},
\bauthor{\bsnm{Burns}, \binits{A.}},
\bauthor{\bsnm{Bril}, \binits{R.J.}},
\bauthor{\bsnm{Lukkien}, \binits{J.J.}}:
\batitle{Controller area network ({CAN}) schedulability analysis: Refuted,
  revisited and revised}.
\bjtitle{Real-Time Systems}
\bvolume{35},
\bfpage{239}--\blpage{272}
(\byear{2007})
\end{barticle}
\endbibitem

%%% 6
\bibitem[\protect\citeauthoryear{Tindell et~al.}{1994}]{tindell1994extendible}
\begin{barticle}
\bauthor{\bsnm{Tindell}, \binits{K.W.}},
\bauthor{\bsnm{Burns}, \binits{A.}},
\bauthor{\bsnm{Wellings}, \binits{A.J.}}:
\batitle{An extendible approach for analyzing fixed priority hard real-time
  tasks}.
\bjtitle{Real-Time Systems}
\bvolume{6},
\bfpage{133}--\blpage{151}
(\byear{1994})
\end{barticle}
\endbibitem

%%% 7
\bibitem[\protect\citeauthoryear{Guan et~al.}{2007}]{guan2007exact}
\begin{bchapter}
\bauthor{\bsnm{Guan}, \binits{N.}},
\bauthor{\bsnm{Gu}, \binits{Z.}},
\bauthor{\bsnm{Deng}, \binits{Q.}},
\bauthor{\bsnm{Gao}, \binits{S.}},
\bauthor{\bsnm{Yu}, \binits{G.}}:
\bctitle{Exact schedulability analysis for static-priority global
  multiprocessor scheduling using model-checking}.
In: \bbtitle{IFIP International Workshop on Software Technolgies for Embedded
  and Ubiquitous Systems},
pp. \bfpage{263}--\blpage{272}
(\byear{2007}).
\bcomment{Springer}
\end{bchapter}
\endbibitem

%%% 8
\bibitem[\protect\citeauthoryear{Sun and Lipari}{2016}]{sun2016pre}
\begin{barticle}
\bauthor{\bsnm{Sun}, \binits{Y.}},
\bauthor{\bsnm{Lipari}, \binits{G.}}:
\batitle{A pre-order relation for exact schedulability test of sporadic tasks
  on multiprocessor global fixed-priority scheduling}.
\bjtitle{Real-Time Systems}
\bvolume{52},
\bfpage{323}--\blpage{355}
(\byear{2016})
\end{barticle}
\endbibitem

%%% 9
\bibitem[\protect\citeauthoryear{Burmyakov et~al.}{2015}]{burmyakov2015exact}
\begin{bchapter}
\bauthor{\bsnm{Burmyakov}, \binits{A.}},
\bauthor{\bsnm{Bini}, \binits{E.}},
\bauthor{\bsnm{Tovar}, \binits{E.}}:
\bctitle{An exact schedulability test for global {FP} using state space
  pruning}.
In: \bbtitle{Proceedings of the 23rd International Conference on Real Time and
  Networks Systems},
pp. \bfpage{225}--\blpage{234}
(\byear{2015})
\end{bchapter}
\endbibitem

%%% 10
\bibitem[\protect\citeauthoryear{Baker and Cirinei}{2007}]{baker2007brute}
\begin{bchapter}
\bauthor{\bsnm{Baker}, \binits{T.P.}},
\bauthor{\bsnm{Cirinei}, \binits{M.}}:
\bctitle{Brute-force determination of multiprocessor schedulability for sets of
  sporadic hard-deadline tasks}.
In: \bbtitle{International Conference On Principles Of DIstributed Systems},
pp. \bfpage{62}--\blpage{75}
(\byear{2007}).
\bcomment{Springer}
\end{bchapter}
\endbibitem

%%% 11
\bibitem[\protect\citeauthoryear{Bonifaci and
  Marchetti-Spaccamela}{2012}]{bonifaci2012feasibility}
\begin{barticle}
\bauthor{\bsnm{Bonifaci}, \binits{V.}},
\bauthor{\bsnm{Marchetti-Spaccamela}, \binits{A.}}:
\batitle{Feasibility analysis of sporadic real-time multiprocessor task
  systems}.
\bjtitle{Algorithmica}
\bvolume{63},
\bfpage{763}--\blpage{780}
(\byear{2012})
\end{barticle}
\endbibitem

%%% 12
\bibitem[\protect\citeauthoryear{Nasri and Brandenburg}{2017}]{SANS}
\begin{bchapter}
\bauthor{\bsnm{Nasri}, \binits{M.}},
\bauthor{\bsnm{Brandenburg}, \binits{B.B.}}:
\bctitle{An exact and sustainable analysis of non-preemptive scheduling}.
In: \bbtitle{2017 IEEE Real-Time Systems Symposium (RTSS)},
pp. \bfpage{12}--\blpage{23}
(\byear{2017}).
\bcomment{IEEE}
\end{bchapter}
\endbibitem

%%% 13
\bibitem[\protect\citeauthoryear{Nasri et~al.}{2018}]{nasri2018response}
\begin{bchapter}
\bauthor{\bsnm{Nasri}, \binits{M.}},
\bauthor{\bsnm{Nelissen}, \binits{G.}},
\bauthor{\bsnm{Brandenburg}, \binits{B.B.}}:
\bctitle{A response-time analysis for non-preemptive job sets under global
  scheduling}.
In: \bbtitle{30th Euromicro Conference on Real-Time Systems},
pp. \bfpage{9}--\blpage{1}
(\byear{2018})
\end{bchapter}
\endbibitem

%%% 14
\bibitem[\protect\citeauthoryear{Nasri et~al.}{2019}]{nasri2019response}
\begin{bchapter}
\bauthor{\bsnm{Nasri}, \binits{M.}},
\bauthor{\bsnm{Nelissen}, \binits{G.}},
\bauthor{\bsnm{Brandenburg}, \binits{B.B.}}:
\bctitle{Response-time analysis of limited-preemptive parallel {DAG} tasks
  under global scheduling}.
In: \bbtitle{31st Conference on Real-Time Systems},
pp. \bfpage{21}--\blpage{1}
(\byear{2019})
\end{bchapter}
\endbibitem

%%% 15
\bibitem[\protect\citeauthoryear{Nogd et~al.}{2020}]{nogd2020response}
\begin{bchapter}
\bauthor{\bsnm{Nogd}, \binits{S.}},
\bauthor{\bsnm{Nelissen}, \binits{G.}},
\bauthor{\bsnm{Nasri}, \binits{M.}},
\bauthor{\bsnm{Brandenburg}, \binits{B.B.}}:
\bctitle{Response-time analysis for non-preemptive global scheduling with
  {FIFO} spin locks}.
In: \bbtitle{2020 IEEE Real-Time Systems Symposium (RTSS)},
pp. \bfpage{115}--\blpage{127}
(\byear{2020}).
\bcomment{IEEE}
\end{bchapter}
\endbibitem

%%% 16
\bibitem[\protect\citeauthoryear{Ranjha et~al.}{2022}]{ranjha2022partial}
\begin{bchapter}
\bauthor{\bsnm{Ranjha}, \binits{S.}},
\bauthor{\bsnm{Nelissen}, \binits{G.}},
\bauthor{\bsnm{Nasri}, \binits{M.}}:
\bctitle{Partial-order reduction for schedule-abstraction-based response-time
  analyses of non-preemptive tasks}.
In: \bbtitle{2022 Real-Time and Embedded Technology and Applications Symposium
  (RTAS)}
(\byear{2022})
\end{bchapter}
\endbibitem

%%% 17
\bibitem[\protect\citeauthoryear{Ranjha et~al.}{2023}]{ranjha2023partial}
\begin{botherref}
\oauthor{\bsnm{Ranjha}, \binits{S.}},
\oauthor{\bsnm{Gohari}, \binits{P.}},
\oauthor{\bsnm{Nelissen}, \binits{G.}},
\oauthor{\bsnm{Nasri}, \binits{M.}}:
Partial-order reduction in reachability-based response-time analyses of
  limited-preemptive {DAG} tasks.
Real-Time Systems,
1--55
(2023)
\end{botherref}
\endbibitem

%%% 18
\bibitem[\protect\citeauthoryear{Nasri and Kargahi}{2014}]{PRM}
\begin{barticle}
\bauthor{\bsnm{Nasri}, \binits{M.}},
\bauthor{\bsnm{Kargahi}, \binits{M.}}:
\batitle{Precautious-{RM}: a predictable non-preemptive scheduling algorithm
  for harmonic tasks}.
\bjtitle{Real-Time Systems}
\bvolume{50},
\bfpage{548}--\blpage{584}
(\byear{2014})
\end{barticle}
\endbibitem

%%% 19
\bibitem[\protect\citeauthoryear{Nasri and Fohler}{2016}]{nasri2016non}
\begin{bchapter}
\bauthor{\bsnm{Nasri}, \binits{M.}},
\bauthor{\bsnm{Fohler}, \binits{G.}}:
\bctitle{Non-work-conserving non-preemptive scheduling: motivations,
  challenges, and potential solutions}.
In: \bbtitle{2016 28th Euromicro Conference on Real-Time Systems (ECRTS)},
pp. \bfpage{165}--\blpage{175}
(\byear{2016}).
\bcomment{IEEE}
\end{bchapter}
\endbibitem

%%% 20
\bibitem[\protect\citeauthoryear{D{\'\i}az et~al.}{2002}]{LEVEL}
\begin{barticle}
\bauthor{\bsnm{D{\'\i}az}, \binits{J.}},
\bauthor{\bsnm{Petit}, \binits{J.}},
\bauthor{\bsnm{Serna}, \binits{M.}}:
\batitle{A survey of graph layout problems}.
\bjtitle{ACM Computing Surveys (CSUR)}
\bvolume{34}(\bissue{3}),
\bfpage{313}--\blpage{356}
(\byear{2002})
\end{barticle}
\endbibitem

%%% 21
\bibitem[\protect\citeauthoryear{Jaro{\v{s}}}{2022}]{jaros2022combination}
\begin{botherref}
\oauthor{\bsnm{Jaro{\v{s}}}, \binits{M.}}:
Combination of time-triggered and event-triggered scheduling.
Master's thesis,
Czech Technical University in Prague
(2022)
\end{botherref}
\endbibitem

%%% 22
\bibitem[\protect\citeauthoryear{Hala{\v{s}}ka}{2023}]{halaska2023combination}
\begin{botherref}
\oauthor{\bsnm{Hala{\v{s}}ka}, \binits{L.}}:
Combination of time-triggered and event-triggered scheduling with dedicated
  resources and precedences.
Master's thesis,
Czech Technical University in Prague
(2023)
\end{botherref}
\endbibitem

\end{thebibliography}

\end{document}